\documentclass[onecolumn,showpacs,preprintnumbers,amsmath,amssymb,pre]{revtex4-1}

\usepackage{color}    
\usepackage{graphicx}
\usepackage{dcolumn}
\usepackage{bm}
\usepackage{subfigure}
\usepackage{amssymb}
\usepackage{multirow}
\usepackage{natbib}
\usepackage{amsmath}
\usepackage{mathrsfs,amsmath}
\graphicspath{{plots/}}
\usepackage{hyperref}
\hypersetup{
    colorlinks = true,
    citecolor = blue
}
\usepackage[utf8]{inputenc}
\usepackage[T1]{fontenc}
\usepackage[normalem]{ulem}
\usepackage{dirtytalk}
\usepackage{appendix}
\usepackage{tabularx}

\renewcommand{\vec}[1]{\mathbf{#1}}
\begin{document}

\title{Force Fields for Molecular Dynamics Simulations of Charged Dust Particles with Finite Size in Complex Plasmas}
\author{N.~Kh.~Bastykova$^{1, 2}$}
\author{N.~E.~Djienbekov$^{1, 2}$}
\author{T.~S.~Ramazanov$^{1, 2}$}
\author{S.~K.~Kodanova$^{1, 2}$}
\email[]{kodanova@physics.kz }

 \affiliation{
 $^1$Institute of Applied Sciences and IT, 280 Baizakov str., 050040 Almaty, Kazakhstan
}
\affiliation{
$^2$Institute for Experimental and Theoretical Physics, Al-Farabi Kazakh National University, 71 Al-Farabi ave.,
 050040 Almaty, Kazakhstan}

\begin{abstract}
To exclude collisions leading to the overlap of finite-sized charged particles in molecular dynamics (MD) simulations in systems like complex (dusty) plasmas, we developed a scheme to generate a pair interaction force functionally depending on the pair distribution of particles. This ensures that the pair distribution function drops to zero at distances between particles corresponding to the point where the surfaces of two particles come into contact. The presented model has a wide utility for the description of various processes in dusty plasmas. As an example, using the non-equilibrium MD method, we performed calculations of the shear viscosity and thermal conductivity of charged spherical particles interacting via screened Yukawa potential.  The role of finite particle size was found to be particularly important for small values of the coupling parameter, i.e., for gas-like and weakly correlated liquid-like systems. Furthermore, we have analyzed the shear viscosity and thermal conductivity in the presence of an external uniform magnetic field.

\end{abstract}
\maketitle


\section{Introduction}\label{sec1}

Research in fields such as dusty plasmas \cite{Wong2018, Petersen2022, FORTOV20051} and the physics of colloids \cite{LIKOS2001267, Stradner2004}  focuses on various systems of strongly correlated classical particles. 
In particular, two-dimensional (2D) systems have recently attracted substantial interest due to their unique structural, dynamic, and transport properties, which differ significantly from those of three-dimensional systems. 
The structural and transport properties of such systems are usually simulated using the molecular dynamics (MD) method. 
For example, using Yukawa pair interaction potential (i.e. a screened Coulomb potential), an extensive investigation of the equilibrium properties and phase diagram of two-dimensional and three-dimensional systems were performed  \cite{Hartmann2005, Vaulina2015, Hamaguchi1997}. 
Considering the system of charged particles interacting via Yukawa potnetial, in Refs. \cite{Khrustalyov2011, Liu2005},  diffusivity and conductivity were studied using the Green-Kubo relations and the equilibrium MD method for various plasma parameters.  Besides of that, the so-called non-equilibrium MD method has also been used to compute transport coefficients of strongly correlated systems \cite{PhysRevE.69.016405, 10.1063/1.4922908, PhysRevE.106.065203}. This approach also allows one to simulate non-equilibrium steady states created in dusty plasma experiments \cite{PhysRevLett.109.185002}. Besides of being a good approximation for the pair-interaction potential in dusty plasmas, the Yukawa-type potentials also emerge naturally in other types of non-ideal plasmas such as warm dense matter \cite{10.1063/1.4932051, Moldabekov_cpp_2017, Moldabekov_cpp_2012}. 
This aspect allows to consider dusty plasmas as an experimental model system for studying various properties such as, e.g, charged particle energy loss \cite{Schwabe_2020, PhysRevE.73.016404, Shukla_2007} on kinetic level to gain a better understanding of similar effects in other types of plasmas \cite{10.1016/j.mre.2017.07.005, Issanova_Meister_2016, 10.1063/1.4829042}.

Furthermore, dusty plasmas were used to investigate the effect of strong magnetic fields on transport and structural properties of charged systems  \cite{PhysRevLett.111.155002, 10.1063/1.4914089, 8688665, Ludwig2018}. For example, magnetoplasmons were observed in strongly correlated system of charged dust particles in the experiment by Hartmann \textit{et al} \cite{PhysRevLett.111.155002}. Such magnetoplasmons in non-ideal plasmas were earlier predicted by computing the density and current fluctuations spectra using data from equilibrium MD simulations. In one of our recent works, we explored the effect of a magnetic field on the thermal conductivity of 2D Yukawa systems using the non-equilibrium molecular dynamics method, revealing that an external uniform magnetic field reduces thermal conductivity, with the effect being weaker at larger coupling parameters \cite{Djienbekov2}.

Being highly useful, such models as Yukawa potential do not take into account the fact that dust particles have finite size, which can be comparable with the mean distance between dust particles.  There are many works devoted to the study of the effect of particle size on the properties of two-dimensional systems, usually by modifying the screening length $\lambda_s$ in the Yukawa potential (e.g., see Refs. \cite{Liu_2007, yang2020dynamics}) or by considering higher-order terms in the multipole expansion of the electrostatic field of dust particles emerging due to their finite size \cite{PhysRevLett.75.4409, 5510173, PhysRevE.93.053204, PhysRevE.102.033205,  boudec2024cartesiansphericalmultipoleexpansions}.
The effect of the finite size of dust particles depends on the density, coupling parameter, and other parameters of plasmas \cite{Liu_2007, yang2020dynamics, 10.1063/1.4887009, Aldakulov_2020, Sundar_2020, PhysRevE.99.063202}.
In previous molecular dynamics simulations, the used interaction models took into account modifications in the electrostatic interaction potential due to the finite size of dust particles. Still, they did not address the constraint that particles cannot overlap. This constraint is a mechanical factor that essentially limits the space available for the dust particles during their movement. In this work, we develop a force field model that effectively incorporates hard collisions of particles into MD simulation. This method allows the use of any electrostatic pair interaction potential in MD simulations while preventing particle overlap. As an example of the application of this scheme, the method is used to calculate the structural properties,  viscosity, and thermal conductivity of a 2D system of charged hard spherical particles interacting via screened Yukawa-type pair potential. We compute the shear viscosity and thermal conductivity in the presence of a strong external uniform magnetic field. The latter provides new insights into the transport properties of the 2D system of charged particles in conditions where the Larmor radius of the particles $r_L$, the radius of the dust gains $r_d$, the mean interparticle distance $a$, and the screening length $\lambda_s$ are of the same order, i.e., $r_L\sim r_d\sim a\sim \lambda_s$. In this regard, we mention the so-called rotating dusty plasma devise that uses the equivalence of the magnetic Lorentz force and the Coriolis inertial force to observe the system of highly magnetized charged dust particles. Therefore, in these types of experiments,  dust particles approach each other at close distances due to Larmor rotation, even if the repulsion between particles is substantial. This means that the simulations need to take into account the exclusion of the configurations containing overlapping particles. 


The parameters at which complex plasmas are realized in nature and the laboratory are highly diverse. In complex plasmas, where the coupling between particles is weak, it is crucial to accurately account for close collisions between particles. Colloidal systems are another area where the size of particles (interacting via electrostatic forces) is comparable to the mean interparticle distance \cite{Vermant_2005, RevModPhys.96.045003, Anderson2002}. The approach presented here can be applied to systems with any electrostatic interaction between particles. As a result, the workflow has general applicability in the fields discussed above. 

In Sec. \ref{s:pot}, we describe the workflow for generating the force field for performing MD simulations. The results for structural and transport properties of the 2D system of spherical charged particles are presented in Sec. \ref{s:no_mag}, where we consider cases with and without magnetic field.  We conclude the paper by summarising the main findings and providing an outlook for the future applications of the proposed force field model.

\section{Force field model for 2D system of charged particles}\label{s:pot}

     \begin{figure} 
    \includegraphics[width=0.35\textwidth]{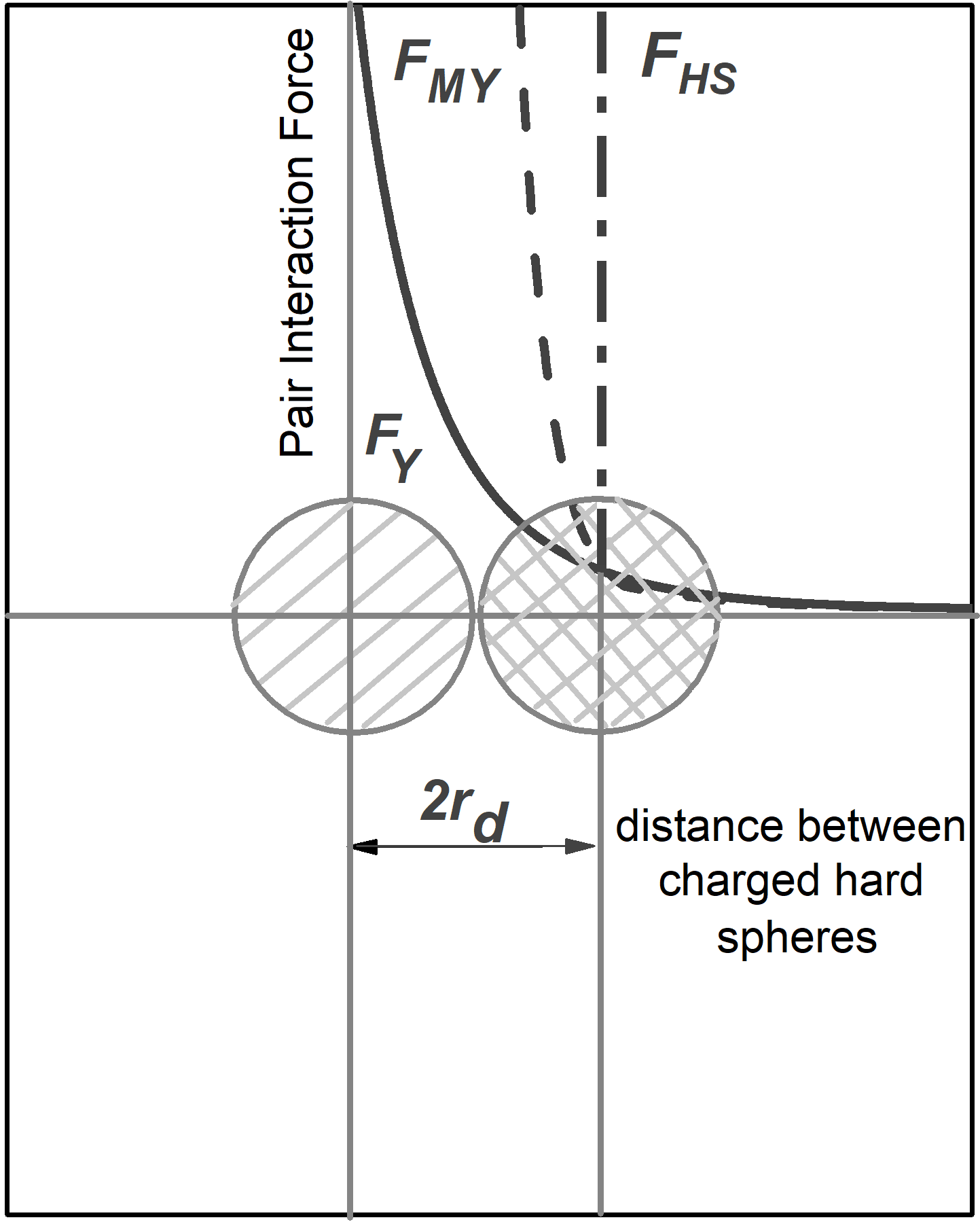}
        \caption{Illustration of the interaction force between two spherical particles as considered in this work.
    \label{fig:0}}
    \end{figure}

We consider charged dust particles that are assumed to be impenetrable during pair collisions. This is a standard picture of dusty plasmas created in low-pressure gas discharges, which is routinely realized in experiments to investigate the physics of strongly correlated systems on the kinetic level \cite{Bonitz_2010}.  In statistical physics, a close-contact collision of two particles of finite size is often treated using the model of hard spheres.
The exclusion of the configurations of positions of the particles from the configurational integral (as defined in statistical physics) that has overlapping particles can generate short-ranged correlations typical for fluids and gases (e.g., see Ref. \cite{Malijevský2008}).
In statistical physics, the correlation between particles refers to the extent to which the properties or states of one particle are related to or dependent on those of another particle within a system. Essentially, correlations emerge from interactions between particles, such as forces. 
The force of the interaction between charged hard spheres $\vec F_{\rm HS}$ tends to positive infinity when the distance between particles is smaller than the diameter of particles $r\leq 2r_d$. This situation during the collision of particles is illustrated in Fig. \ref{fig:0} by a vertical dash-doted line. In illustration \ref{fig:0}, we also depict the force of the Yukawa potential $\vec F_{Y}$, commonly used in MD simulations of charged dust particles \cite{FORTOV20051, Bonitz_2010, PhysRevResearch.3.043187}. The success of the Yukawa potential-based models for the simulation of the properties of dusty plasmas is due to the fact that they capture the key feature of the interaction of charged particles in plasmas: screening due to the polarization of surrounding plasmas around charged grains. To account for the fact that dust particles cannot penetrate each other [we do not consider the destruction of dust particles], one can use $\vec F_{Y}+\vec F_{\rm HS}$ as a force of pair interaction.
Besides the Yukawa potential, one can combine any other types of pair interaction force (potential) with $\vec F_{Y}$. A more general form of the electrostatic interaction potential $\vec F_{\rm E}$ can take into account such effects as non-linear screening \cite{Tsytovich2013, PhysRevLett.97.258302}, higher-order multipole moments \cite{PhysRevE.93.053204, boudec2024cartesiansphericalmultipoleexpansions},  wakefield due to a stationary flux of ions \cite{Ludwig2018, 5510173}, etc. 
Thus, in general, one can use $\vec F_{\rm E}+\vec F_{\rm HS}$ to take into account that particles cannot penetrate each other and, in this way, consistently reduce the possible number particles configurations.
Being a simple model for the approximation of the close collision of dust particles, it has a disadvantage 
from a numerical point of view. Namely, it is, to a certain degree, challenging to accurately model the contact collision of hard spheres in MD simulations. One can treat such collisions as elastic scattering of particles. For this, a small enough time step is required to avoid the overlapping of particles during the discrete shifts of their coordinates in the MD simulations. The length of the time step $\tau$ allowing effectively treat such elastic collisions due to contact of hard spheres will depend on the system parameters. It is not clear in advance what $\tau$ value one needs to choose for this purpose. This becomes more complicated in the presence of a strong external magnetic field, due to which particles rotate with the characteristic Larmor radius comparable with the mean distance between particles. As an alternative, one can use a strong repulsive force at close distances between particles that effectively eliminates the overlapping of particles during MD simulation. 
The parameters of such short-range strong repulsion force $\vec F_{\rm SR}$ will also depend on plasma characteristics, but as we show here, one can find the parameterization of $\vec F_{\rm SR}$ that is independent of the strength of the external magnetic field. Consequently, for MD simulations, we use  $\vec F_{\rm SR}+\vec F_{\rm E}$ (see the dashed line in the illustration \ref{fig:0}, where we depict a modified Yukawa force of pair interaction $\vec F_{\rm MY}=\vec F_{\rm SR}+\vec F_{\rm Y}$ ).
This allows us to efficiently simulate charged dust particles with finite size at different external magnetic field induction strength values. The parameters of $\vec F_{\rm SR}$ are chosen in such a way that the probability of the event where particles overlap tends to zero.

We consider the pair interaction force between particles in the form: 
\begin{equation}\label{eq:1}
    \vec F(\vec r)=\begin{cases}
    &\vec F_{\rm SR}(\vec r)  ~\left[ r\leq 2b_{\rm eff}\right] \\
   & \vec F_{\rm E}(\vec r) ~\left[ r> 2b_{\rm eff}\right],
\end{cases} 
\end{equation}
where $b_{\rm eff}\geq r_d$ defines the distances $r\leq 2b_{\rm eff}$ at which short-range repulsion is activated.
At $r\geq 2b_{\rm eff}$, the pair interaction force is given by the electrostatic interaction $\vec F(\vec r)=\vec F_{\rm E}(\vec r)$.

For the repulsion at short ranges, we use
\begin{equation}\label{eq:2}
    \vec F_{\rm SR}(\vec r)=-\vec\nabla \left(\frac{C\left[b_{\rm eff}\right]}{r^{\alpha}}\right), 
\end{equation}
with $\alpha\gg 1$ and the parameter $C\left[b_{\rm eff}\right]$ being defined from the stitching (continuity) condition:
\begin{equation}\label{eq:3}
    \left.\vec F_{\rm SR}(\vec r) \right|_{ r=2b_{\rm eff}} {\equiv} \left. \vec F_{\rm E}(\vec r)\right|_{ r=2b_{\rm eff}}.
\end{equation}


\begin{figure*}[t!]
    \centering
    \begin{minipage}{.45\textwidth}
        \includegraphics[width=\linewidth]{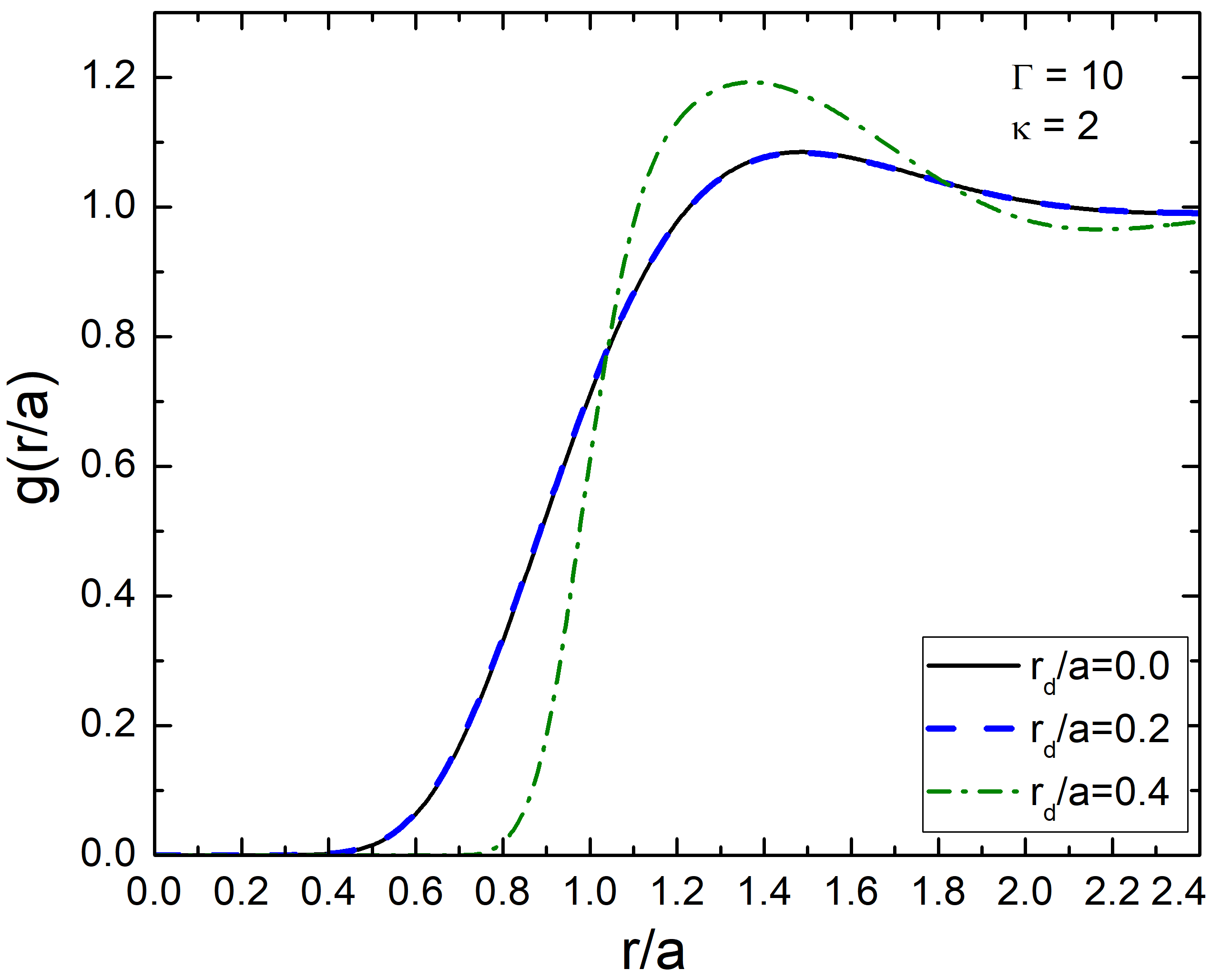}
        \centering
        \textbf{a)} 
    \end{minipage}%
    \hspace{1mm} 
    \begin{minipage}{.45\textwidth}
        \includegraphics[width=\linewidth]{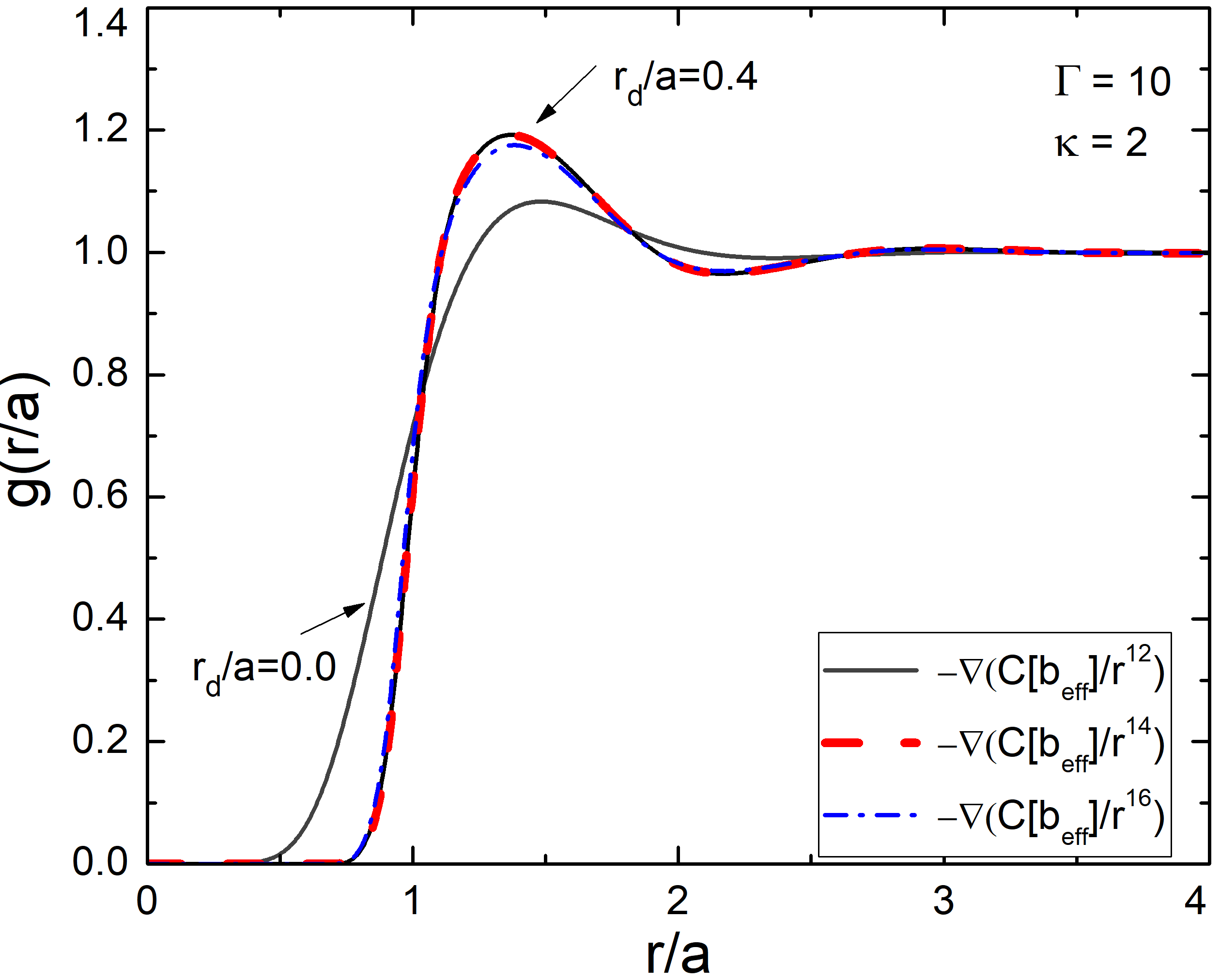}
        \centering
        \textbf{b)} 
    \end{minipage}%
    \caption{ Left: The radial distribution function (RDF) for different values of dust particle's radius $r_{d}/a$. Right: The RDF  dependence on the parameters of the short-ranged repulsion with $r_{d}/a=0.4$ and the parameter $\alpha$ in Eq. (\ref{eq:2}) being varied from 12 to 16.  The RDF values were computed for $\Gamma = 10$ and $\kappa=2$.}
    \label{fig:RDF}
\end{figure*}

For a given electrostatic potential of pair interaction between dust particles, the model defined by Eqs.~(\ref{eq:1})-(\ref{eq:3})
requires a condition determining $b_{\rm eff}$.  As mentioned, the main functionality of $\vec F_{\rm SR}(\vec r)$ is the exclusion of the part of the configurations that have overlapping particles. On a statistical level, we achieve these by employing the radial pair distribution function $g(\vec r)$ (RDF) and the condition:
\begin{equation}\label{eq:condition}
    \left. g(\vec r)\right|_{r\leq 2r_d}=0,
\end{equation}
which means that the probability of finding two particles at the distances $r<2r_d$ smaller than the diameter $2r_d$ of dust particles is zero. 
In practice, in our calculations we require that the RDF is smaller than $\epsilon=10^{-3}$ at $r\leq 2r_d$. 
In this way, we eliminate configurations where the overlapping of a pair of particles occurs.  
Therefore, for a given set of plasma parameters, condition (\ref{eq:condition}) allows us to find $b_{\rm eff}$. We stress that in this way, the pair interaction force of particles $\vec F([g], \vec r)$ is a functional of the RDF. Furthermore, we utilize the fact that an external uniform magnetic field does not affect the RDF of the equilibrium system, meaning that the computed force field is the same for both magnetized and unmagnetized systems. Indeed, a homogeneous magnetic field $\vec B$ does not change the statics and structure of the equilibrium system because the configurational part of the Hamiltonian is independent of $\vec B$. This is numerically illustrated in the appendix.

The presented model can be used with any type of electrostatic pair interaction potential to generate $\vec F_{E}$.
Keeping the discussions on a general level, we use the Yukawa pair interaction potential for the calculation of  $\vec F_{E}$, which yields:
\begin{equation}
     \vec F_{E}(\vec r)=\vec F_{Y}(\vec r)=-k_BTa\,\vec\nabla \left[\frac{\Gamma}{r}\exp\left(-\kappa \frac{r}{a}\right)\right],
     \label{eq:Yukawa}
\end{equation}
where $\kappa$ is the dimensionless screening parameter and $\Gamma=Q^2/k_BTa$ is the coupling parameter of particles with charge $Q$, temperature $T$, and the mean inter particle distance $a$.

     \begin{figure} 
    \includegraphics[width=0.5\textwidth]{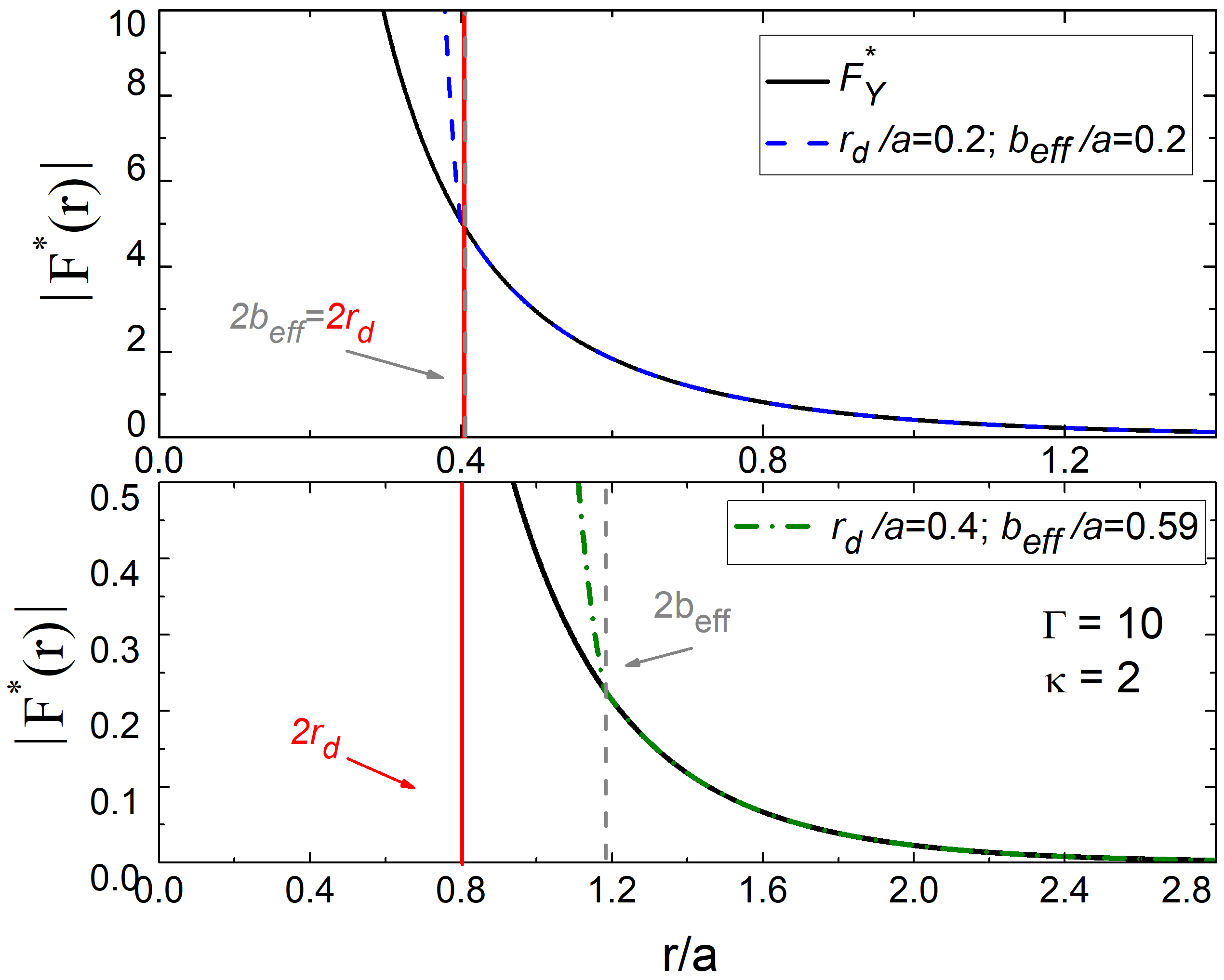}
        \caption{The interaction force magnitude $\left|F^{*}(r)\right|$ in the units of $Q^2/a^2 k_B T)$ as a function of inter particle distance for $r_d=0.2a$ and $r_d=0.4a$ at $\Gamma=10$ and $\kappa=2$. The vertical lines indicate distances at which the short-range repulsion is activated. 
    \label{fig:Force}}
    \end{figure}

Using the force field model (\ref{eq:1}), we demonstrate the effect of short-range repulsion on the RDF of the system of Yukawa particles with a finite size in Fig. \ref{fig:RDF}. In the left panel, we show the RDF computed setting $\alpha=12$ in Eq. (\ref{eq:2}) at $\Gamma=10$ and $\kappa=2$. In this case, we compare the RDF values computed for $r_d=0.2a$ and $r_d=0.4a$ with the data obtained by considering particles as point charges. From the left panel of Fig. \ref{fig:RDF}, we see that the RDF of point charges and the RDF of particles with the radius $r_d=0.2a$ are identical. The reason for this agreement is that the RDF of point charges at $\Gamma=10$ and $\kappa=2$ has the value $g(r)<\epsilon$ at $r=2r_d$ meaning that the particles do not approach each other close enough to lead to a significant overlap effect of particles.
The increase of the dust particle radius to $r_d=0.4a$ results in a significant deviation of the RDF from that of point charges, i.e., the RDF computed by setting $r_d=0$. 
Indeed, the RDF of point charges has a value of about $g(r)\simeq 0.4$ at $r=0.8a$, meaning that a significant correction is needed to avoid the overlap of particles. Introducing such correction using the force field model presented in this work for $r_d=0.4a$ results in the drop of the RDF value to $g(r)\leq \epsilon$  at $r\simeq 0.8a$. This effectively means a stronger correlation between particles, which leads to the increase of the maximum of the RDF from $g_{\rm max}\simeq 1.05$ for point charges to about $g_{\rm max}\simeq 1.2$ for the particles with the finite size of $r_d=0.4a$.
In the right panel of Fig. \ref{fig:RDF}, we compare the RDF values computed using $\alpha=12$, $\alpha=14$, and $\alpha=14$ for $r_d/a=0.4$ at $\Gamma=10$, $\kappa=2$. From the right panel of Fig. \ref{fig:RDF}, we see that further increase of the parameter $\alpha$ from 12 to 13 and 14 does not affect the RDF of particles. 


In Fig. \ref{fig:Force}, we show the pair interaction forces used to compute the RDFs in Fig. \ref{fig:RDF} with $\alpha=12$. In Fig. \ref{fig:Force}, we show the force $\vec F_{Y}$ of the pair interaction of point charges as computed using the Yukawa model. In the top (bottom) panel of Fig. \ref{fig:Force}, we compare the interaction force (\ref{eq:1}) for $r_d=0.2a$ ($r_d=0.4a$) with $\vec F_{Y}$. At  $r_d=0.2a$, the additional short-range repulsion is activated at the distances $r\leq 2r_d=0.4a$. At $r_d=0.4a$, the short-range repulsion is activated at the distances between particles $r\leq 1.18 a$.

\begin{figure} 
    \includegraphics[width=0.5\textwidth]{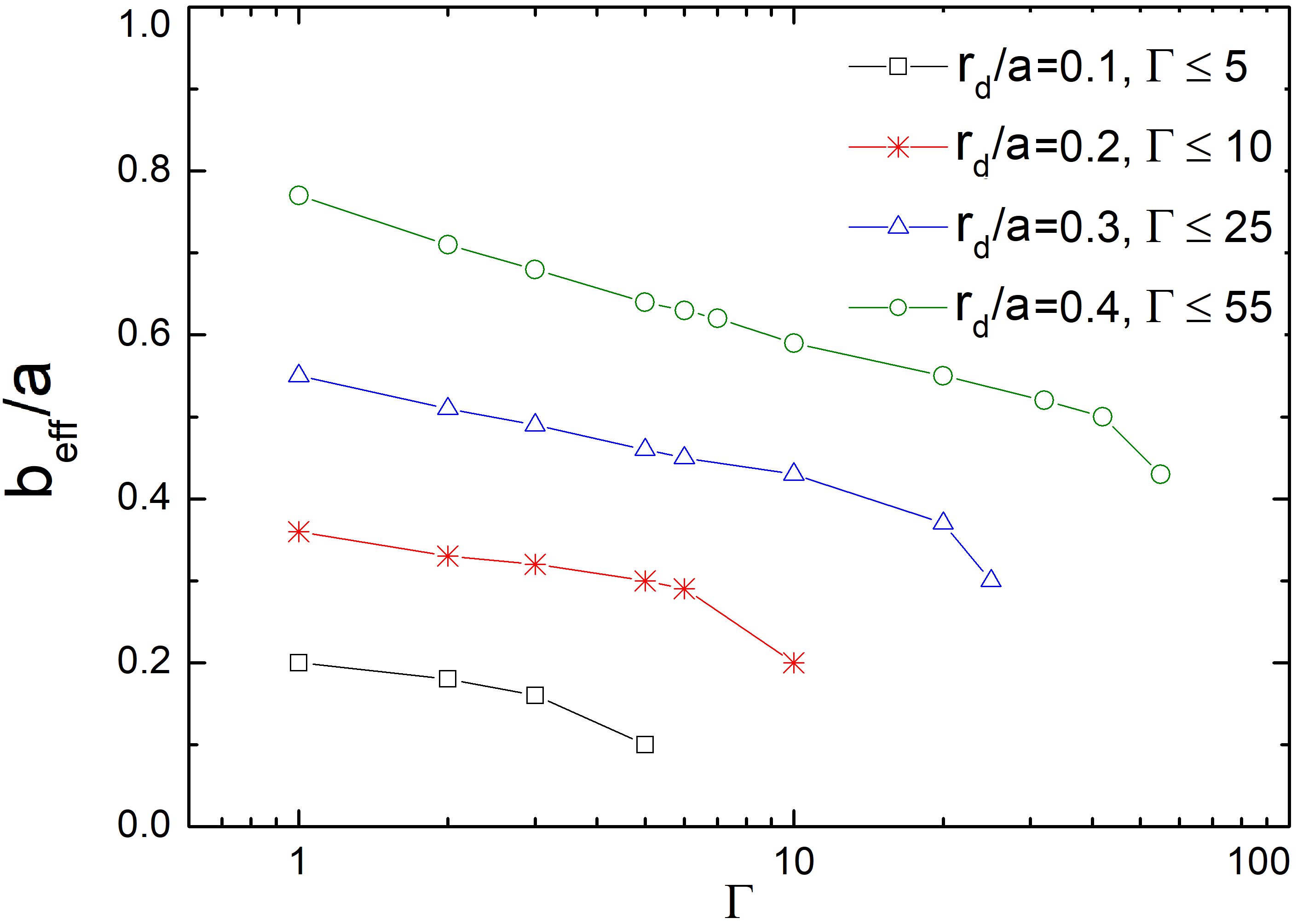}
        \caption{The parameter $b_{\rm eff}$ in Eq. (\ref{eq:1}) defining the range of the additional short-range repulsion. The results are computed for different $\Gamma$ values and different sizes of particles, with the screening parameter being set to $\kappa=2$.
    \label{fig:beff}}
    \end{figure}

\begin{figure}
\includegraphics[width=0.5\textwidth]{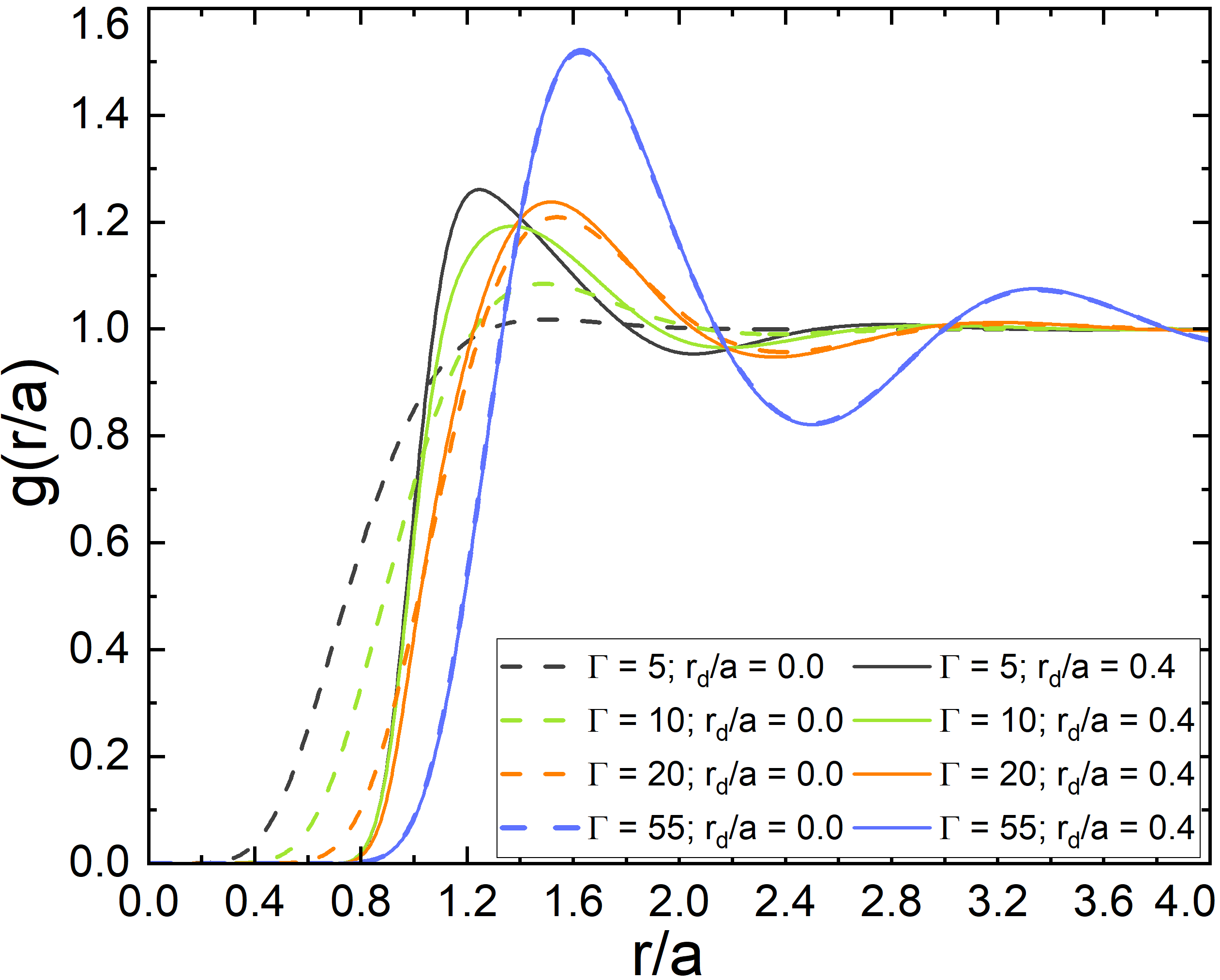}
\caption{The dependence of the radial distribution function (RDF) on the coupling parameter $\Gamma$ for pointlike particles $r_d=0.0a$ (dash lines) and for particles with radii $r_d=0.4a$ (solid lines). 
\label{fig:RDF_Gamma}}
\end{figure}

\begin{figure}
  \centering
    \includegraphics[width=0.5\textwidth]{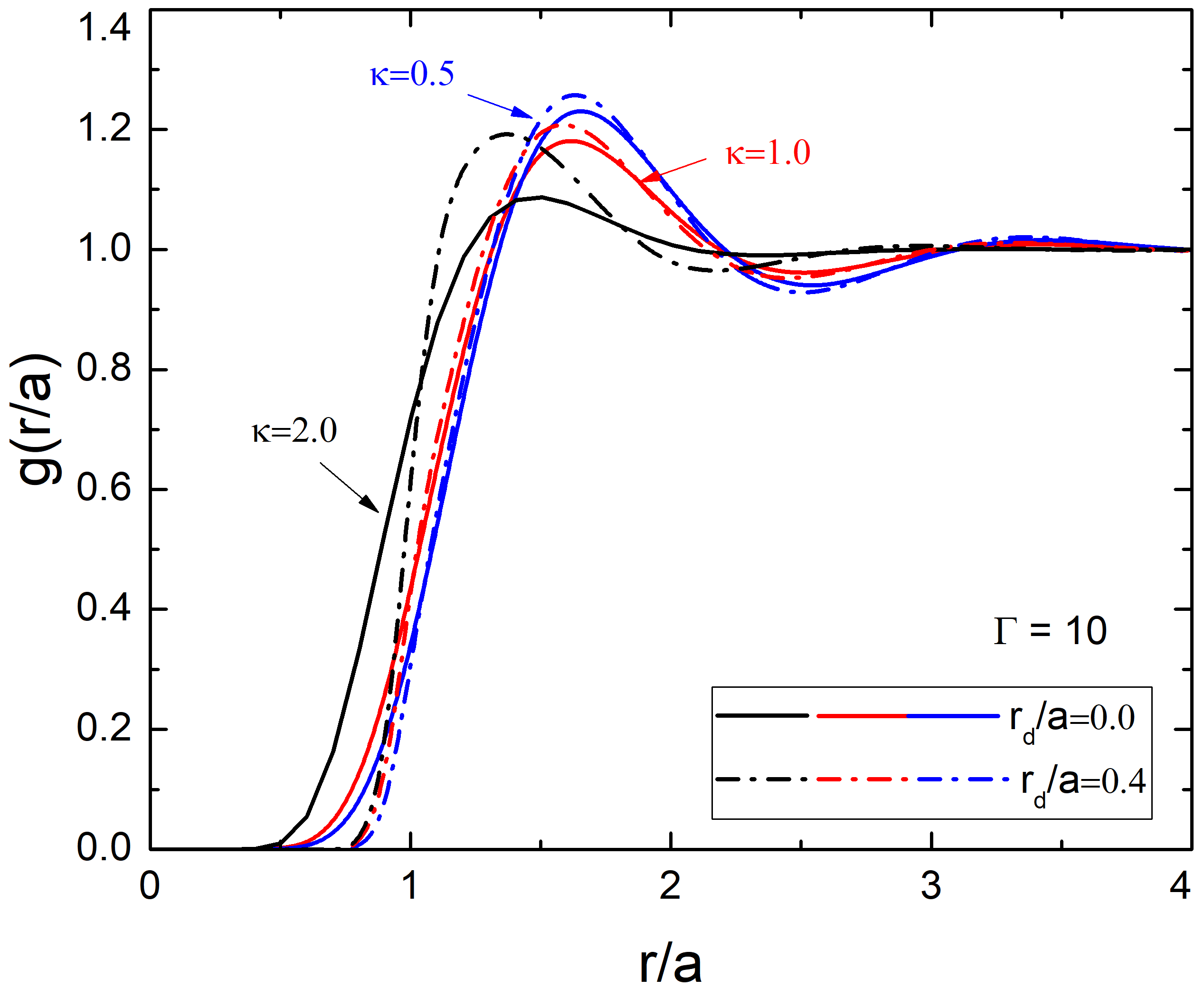}
        \caption{The radial distribution function (RDF) for $\kappa = 2.0, 1.0, 0.5$ at $\Gamma = 10$. The results are presented for  $r_{d}/a=0.0$ and $r_{d}/a=0.4$.}
    \label{fig:17}
\end{figure}

To study the transport properties of the 2D system of charged particles, we recompute the parameter $b_{\rm eff}$ in Eq. (\ref{eq:1}) for different $\Gamma$ values and different sizes of particles with the screening parameter being set to $\kappa=2$.
The values of $b_{\rm eff}$ are shown in Fig. \ref{fig:beff}. The general trend is that $b_{\rm eff}$ decreases with the increase in $\Gamma$. For a given $\kappa$ and the radius of particles $r_d$, after certain values of $\Gamma$, there is no need to use an additional short-range repulsion since the Coulomb repulsion becomes strong enough to keep particles at distances larger than $2r_d$. 
This is illustrated in Fig. \ref{fig:RDF_Gamma}, where we show the RDF at different $\Gamma$ for $r_d=0.4a$ and for the system of charged point charges (i.e., $r_d=0$). From Fig. \ref{fig:RDF_Gamma}, one can see that the difference between RDFs computed for $r_d=0.4a$ and $r_d=0$ decreases with the increase in the coupling parameter $\Gamma$. At $\Gamma=55$, the results for two systems become indistinguishable. Similarly, at a given value of the coupling parameter $\Gamma$, the decrease in the screening parameter $\kappa$ results in the increase in the average distance between particles and reduces close two-particle collision rates. As a result, the importance of the dust particles' finite size effect diminishes with the decrease in $\kappa$. This is illustrated in Fig. \ref{fig:17}, where we show the RDF values computed for $\kappa=2$, $\kappa=1$ and $\kappa=0.5$. From fig. \ref{fig:17}, we clearly see that the RDF of finite-sized particles becomes closer the RDF of point charges with the decrease in the screening parameter. This also means that the effects related to the finite size of dust particles become stronger and more pronounced with the increase in the screening parameter.

\section{Simulation parameters}
The introduced force field model is used to study the transport properties of a 2D system of charged particles without and with an external uniform magnetic field directed normally to the 2D plane of the particles' location. For that,
we consider an ensemble of particles interacting through the force given by Eq. (\ref{eq:1}) and confined in the simulation box with periodic boundary conditions. We set $\alpha=12$ in Eq. (\ref{eq:2}). The Yukawa force (\ref{eq:Yukawa}) is used with $\kappa=2$, which is representative of laboratory dusty plasmas \cite{Bonitz_2010}. To compute transport coefficients, we use both equilibrium and non-equilibrium MD simulation methods.
In the MD simulations, we set the time step $\Delta t=0.01$ (in the units of the inverse plasma frequency of charged particles $\omega_p^{-1}$).
We note that for the considered system, the introduced short-ranged repulsion does not require a substantially smaller MD time step. This is demonstrated in Fig.\ref{fig:15}, where we show a) the dependence of the energy of the system on the MD time step at $0.01\leq \Delta t \omega_p \leq 0.08$ and b) the corresponding RDF values. The energy of the system is given in the units of Coulomb interaction energy of two particles at a mean interparticle distance $\epsilon = Q^2/(4\pi \epsilon_0 a)$. System properties were calculated after the system reached the equilibrium state ($t\omega > 0.0$ in  Fig.\ref{fig:15}). As can be seen, the energies are accurately conserved at all considered simulation time steps. Also, the radial distribution function (RDF) does not change with the variation of the time step.  

\begin{figure*}[ht]
    \centering
    \begin{minipage}{.45\textwidth}
        \includegraphics[width=\linewidth]{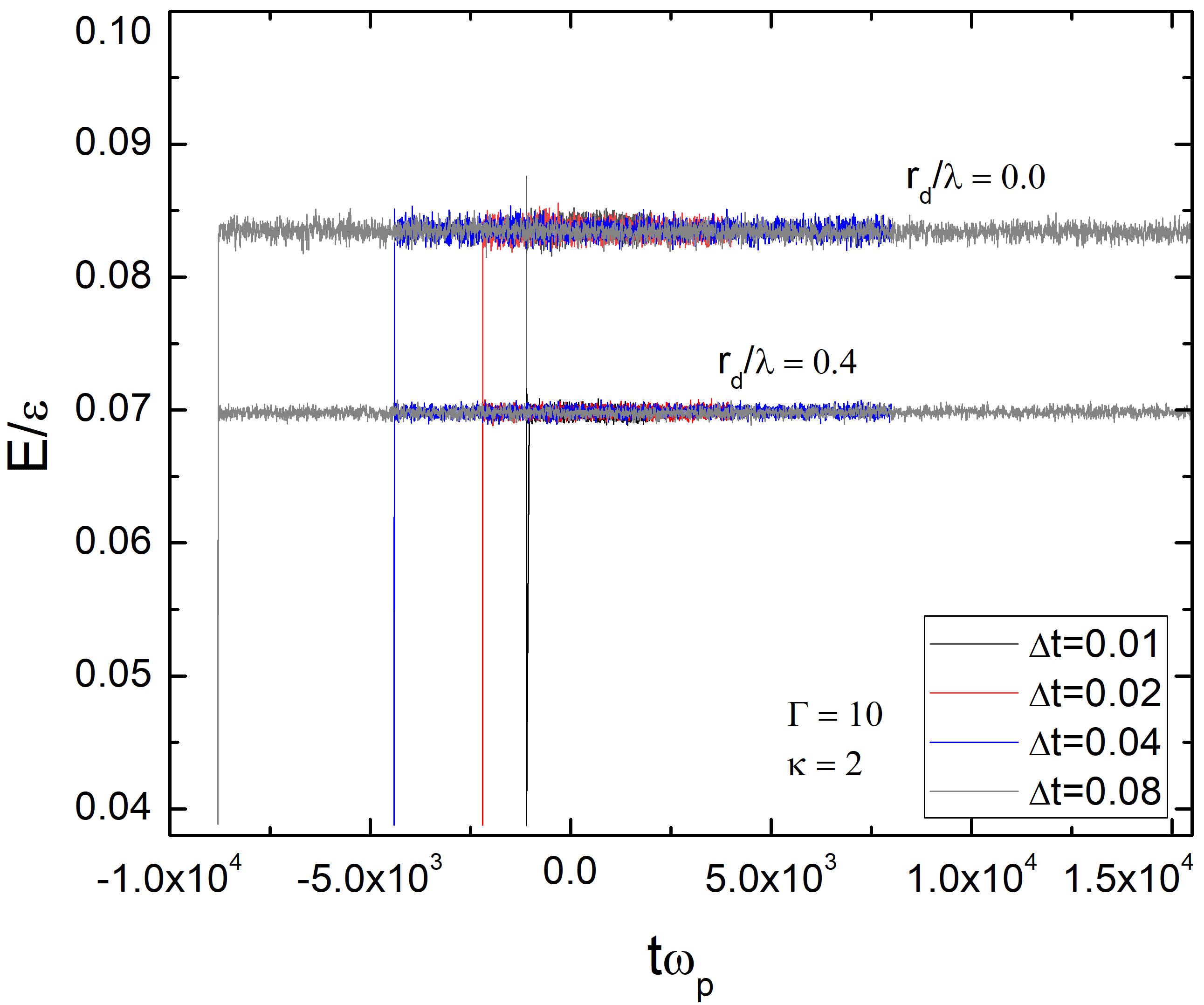}
        \centering
        \textbf{a)} 
    \end{minipage}%
    \hspace{1mm} 
    \begin{minipage}{.45\textwidth}
        \includegraphics[width=\linewidth]{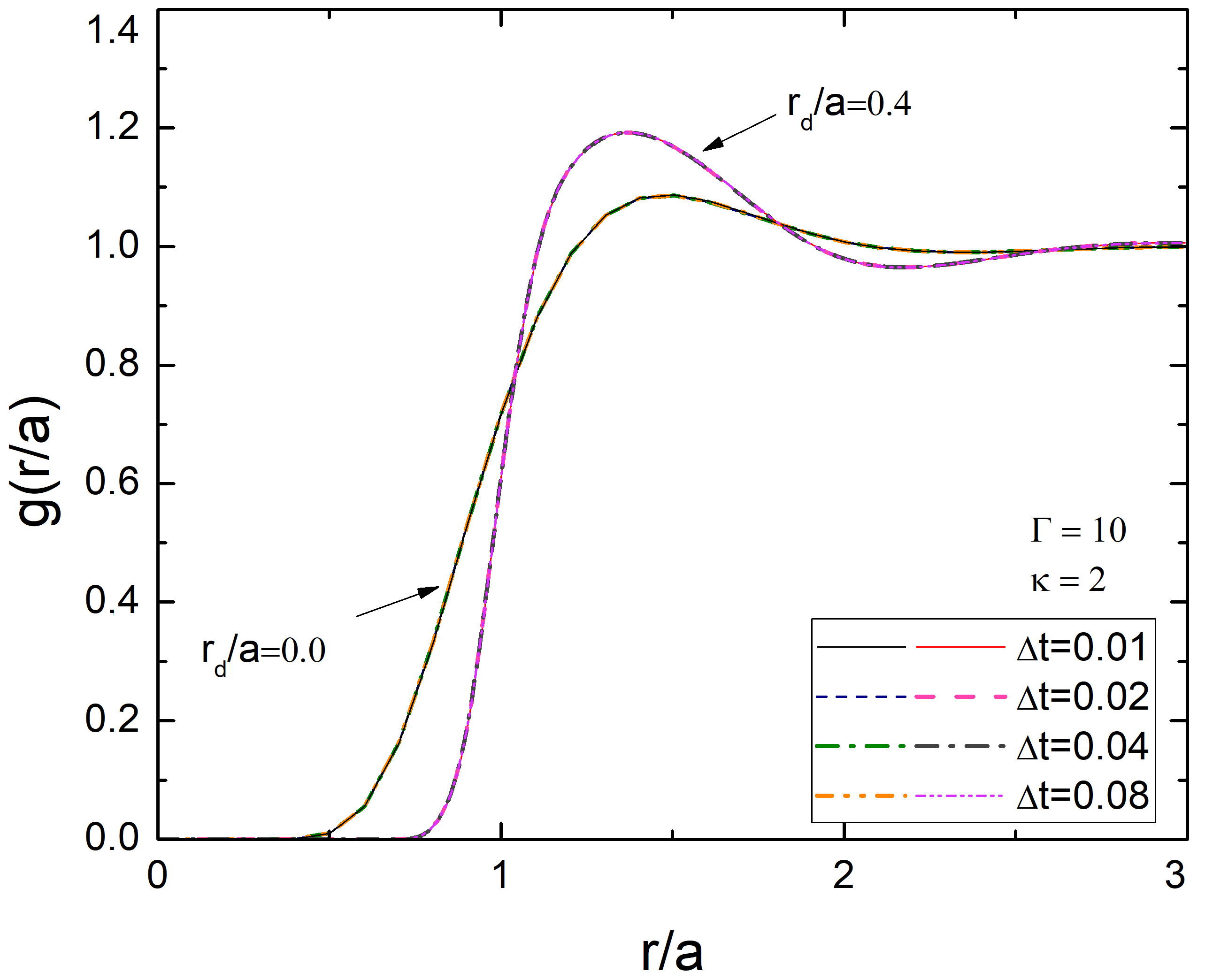}
        \centering
        \textbf{b)} 
    \end{minipage}%
    \caption{  The energy distribution (a) and the radial distribution function (RDF) (b) at different time step of the simulation at $\Gamma = 10$, $\kappa = 2$. The results are presented for different values of the particle radius $r_{d}/a=0.0$ and $r_{d}/a=0.4$.}
    \label{fig:15}
\end{figure*}

The number of particles in the equilibrium MD method is set to $N = 10000$. In the non-equilibrium MD simulations, we set the number of particles to $N = 1600$. The results are averaged over multiple independent simulations to obtain converged results.  To solve the equations of motion of charged particles in an external magnetic field, we used the modified Velocity Verlet algorithm presented in Ref. \cite{Spreiter1999ClassicalMD}. 

The simulation results are given with the length scale normalized by the mean interparticle distance in a 2D system $a=1/\sqrt(\pi n)$ with $n$ being the number density of particles. In this case, the simulation cell size $L$ is defined by the number of particles $L/a=\sqrt{\pi N}$.

\section{Transport properties}\label{s:no_mag}
\subsection{Shear viscosity} \label{sec5a}
The calculation of shear viscosity by the non-equilibrium molecular dynamic (NEMD) method is described in more detail in \cite{Muller,Sanbonmatsu,Hartmann,Djienbekov}. A simulation square box with a side length of $L$ consisting $N$ particles is simulated with periodic boundary conditions. Two horizontal slabs at levels $y=L/4$ and $y=3L/4$ are assigned in the simulation box. In these slabs, the particles with maximal and minimal velocity $v_x$ are selected, and then their moments are interchanged with a certain frequency. Thus, this exchange of particle velocities simulates two currents flowing in opposite directions.

\begin{figure*}[ht]
    \centering
    \begin{minipage}{.32\textwidth}
        \includegraphics[width=\linewidth]{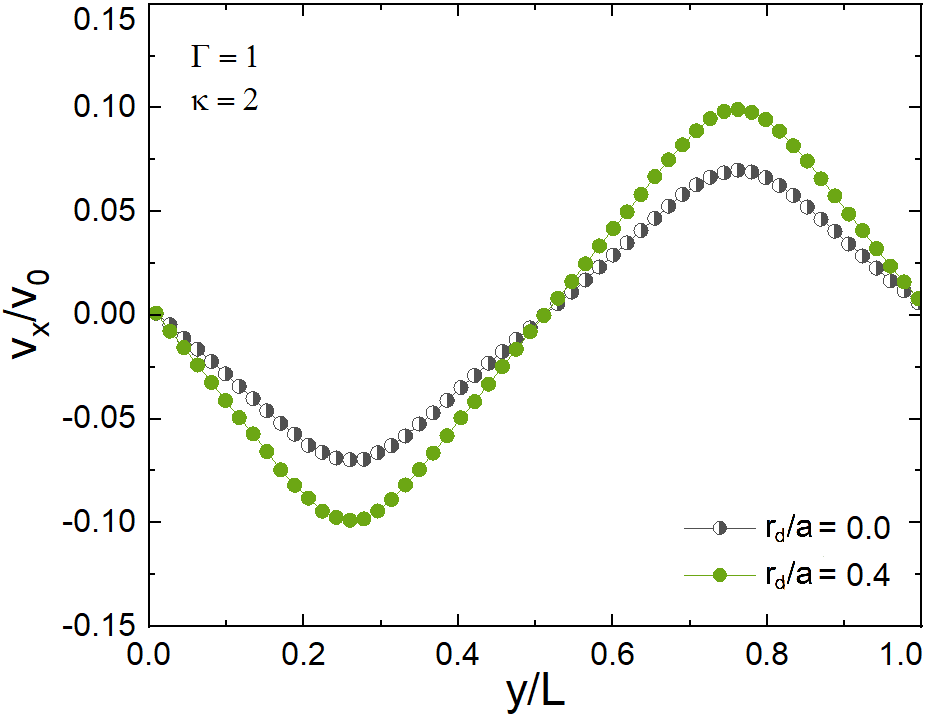}
        \centering
        \textbf{a)} 
    \end{minipage}%
    \hspace{1mm} 
    \begin{minipage}{.32\textwidth}
        \includegraphics[width=\linewidth]{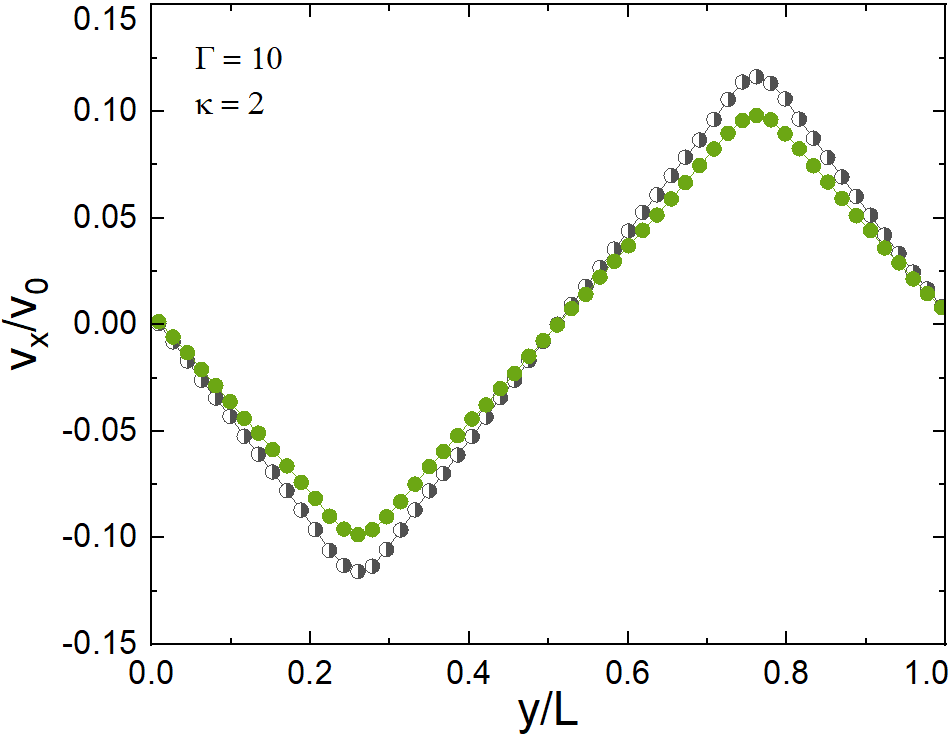}
        \centering
        \textbf{b)} 
    \end{minipage}%
     \hspace{1mm} 
       \begin{minipage}{.32\textwidth}
        \includegraphics[width=\linewidth]{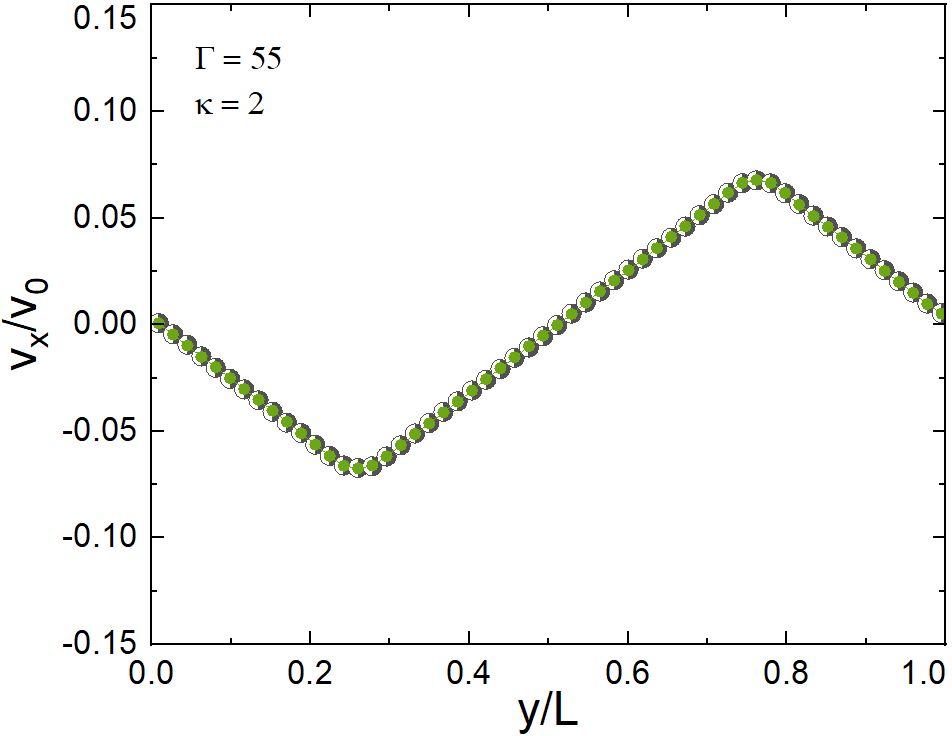}
        \centering
        \textbf{c)} 
    \end{minipage}%
    \caption{Velocity profiles perpendicular to the direction of flows for a)  $\Gamma = 1$, b) $\Gamma = 10$, and c) $\Gamma = 55$. The results are presented for different values of the particle radius $r_{d}/a=0.0$ and $r_{d}/a=0.4$ and without an external magnetic field.}
    \label{fig:7}
\end{figure*}

This generates a non-equilibrium but steady velocity distribution, as shown in Fig \ref{fig:7} for $\Gamma=1$, $\Gamma=10$, and $\Gamma=55$ at $\kappa=2$ without magnetic field. The velocity distribution reaches minimum and maximum values at  $y=L/4$ and $y=3L/4$, respectively.
Between maximum and minimum values, the velocity distribution has linear dependence. This is indicative that the perturbation of the system is within a linear response regime. Therefore, one can write the linear relation between   the momentum flux per unit length $j_x(p_x)=\Delta p/2Lt$  and the shear rate $\partial v_x/\partial y$  induced by two oppositely directed fluxes (in our case along the $x$ axis):
\begin{equation}\label{eq:eta}
    j_x(p_x)=-\eta \,\frac{\partial v_x}{\partial y},
\end{equation}
with the proportionality coefficient being defined as shear viscosity $\eta$. The minus sign takes into account that the velocity gradient and the flux of particles have opposite directions. Relation (\ref{eq:eta}) is used to compute the shear viscosity using the NEMD methods.
We compute the shear viscosity in units of $\eta_0=mn\omega_pa^2$.

\begin{figure}
    \includegraphics[width=0.5\textwidth]{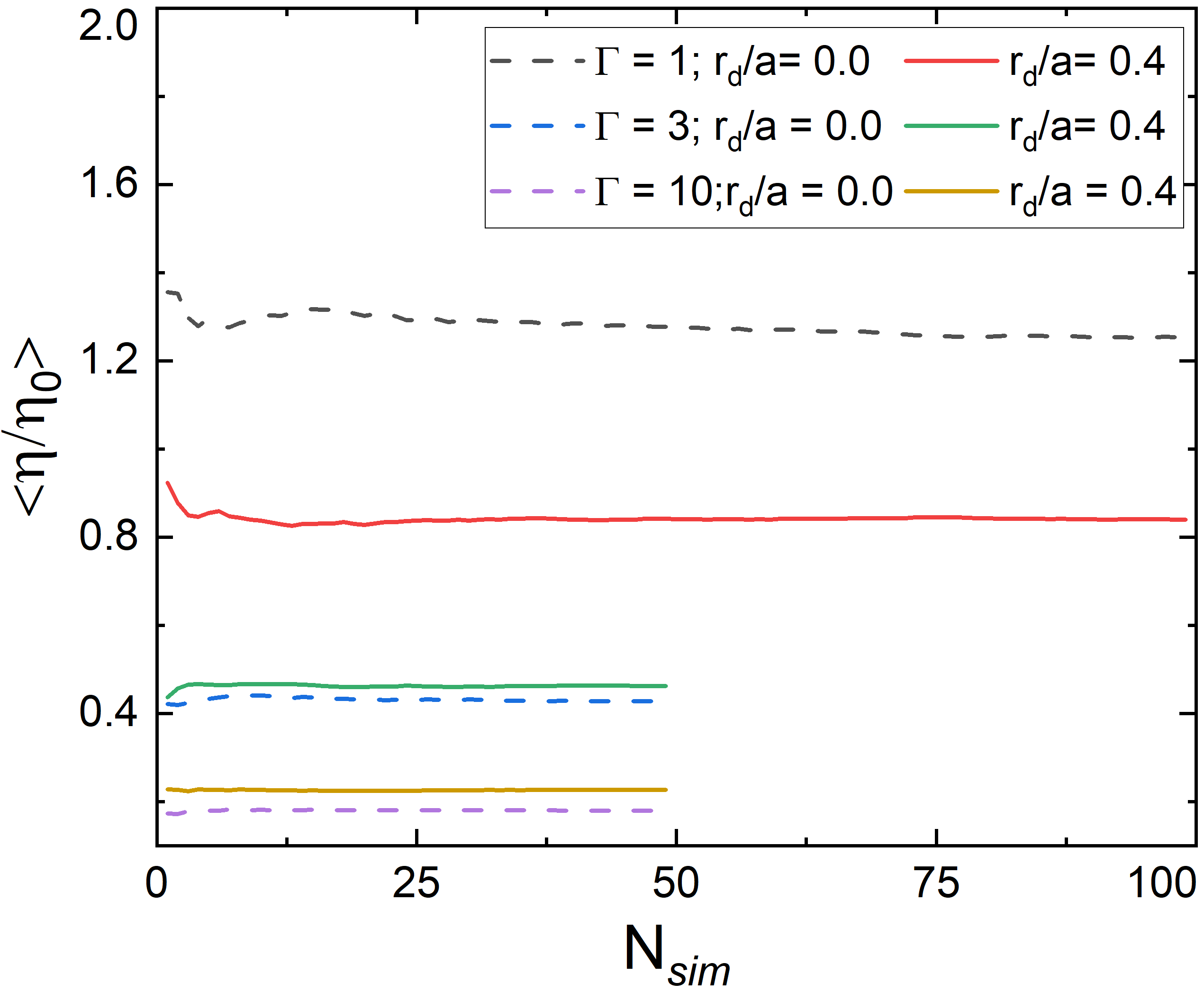}
        \caption{The dependence of the shear viscosity on the number of the used values from independent simulations in the averaging. The results are for $r_d=0.0a$ and $r_d=0.4a$ at $\Gamma=1, 3, 10$. The calculations are performed without an external magnetic field.
        }
    \label{fig:8}
\end{figure}

To obtain converged data, we have averaged the results over a large number of independent NEMD simulations.
For example, for $\Gamma=1$, we averaged over $N_{\rm sim}=100$ independent calculations. In general, the increase in $\Gamma$ reduces the number of needed independent MD runs for averaging. This is illustrated in Fig. \ref{fig:8} for the 2D system of point charges and 2D system of charged particles with radius $r_d=0.4a$ at $\Gamma=1$, $\Gamma=3$, and $\Gamma=10$.  

To cross-check and verify the NEMD implementation, we have computed the shear viscosity using the standard Green-Kubo formula and the data from the equilibrium MD (EMD) simulations. In the EMD method, the shear viscosity is calculated from the Green-Kubo relation using the stress autocorrelation function:
\begin{equation}\label{eq:Green-Kubo}
\eta = \frac{1}{Sk_BT}\int_{0}^{\infty} C(t) \,dt, 
\end{equation}
where $S$ is the area of the simulation box and $C(t)=\left<P^{xy}(t)P^{xy}(0)\right>$ is the stress autocorrelation function (SACF) of particles, $P^{xy}$ is the off-diagonal element of the pressure tensor,
\begin{equation}\label{eq:Stress tensor element}
P^{xy}=\sum_{i=1}^{N}\left[mv_{ix}v_{iy}-\frac{1}{2}\sum_{i\neq j}^{N}\frac{x_{ij}y_{ij}}{r_{ij}}\frac{\partial V(r_{ij})}{\partial r_{ij}}\right],
\end{equation}
where $N$ is the number of particles and $r_{ij} = |\mathbf{r_{i}-r_{j}}|$.

\begin{figure}
    \includegraphics[width=0.5\textwidth]{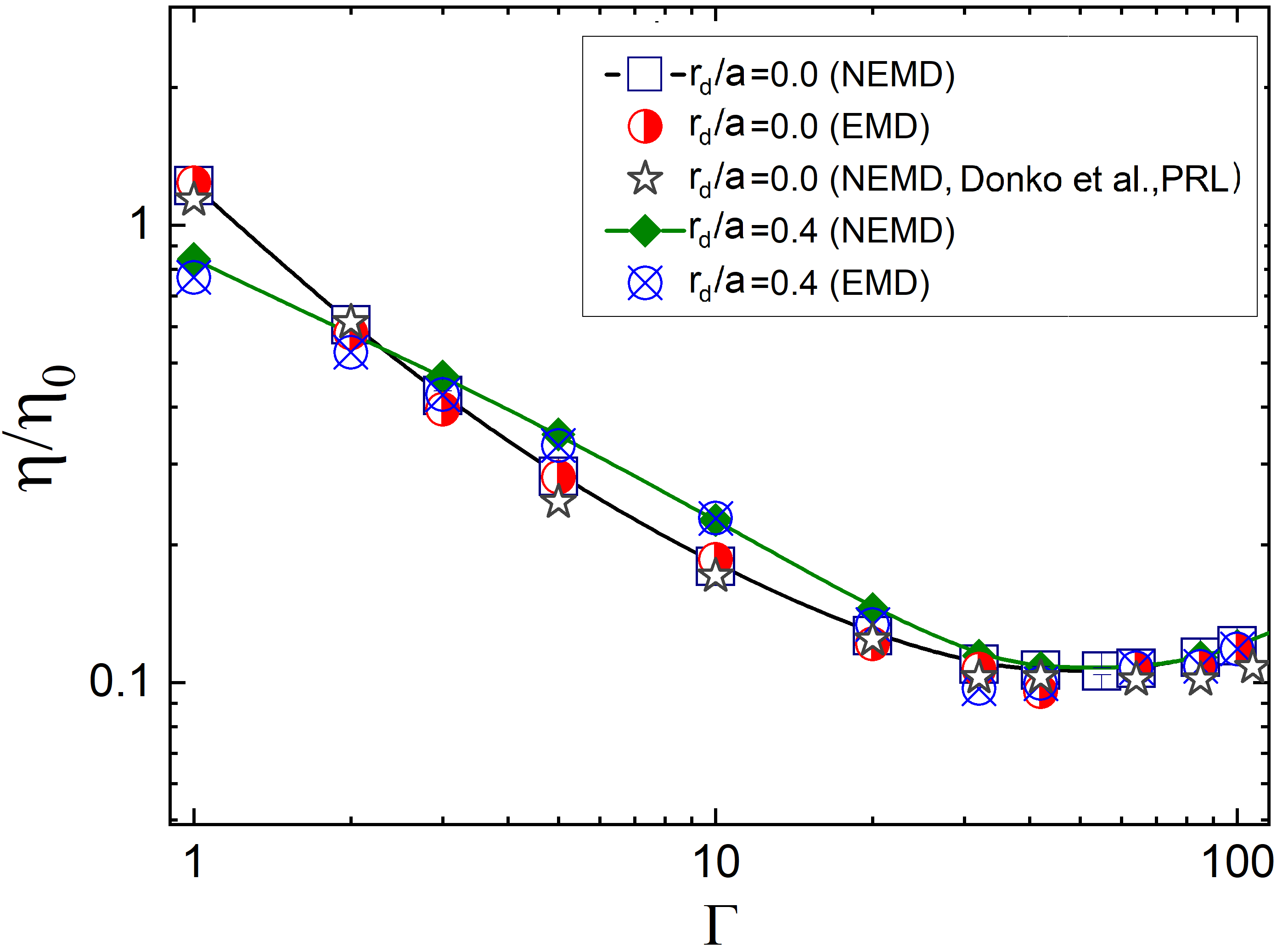}
        \caption{The shear viscosity as a function of the coupling parameter $\Gamma$. The NEMD and EMD calculation results are shown. In addition, we provide a comparison with the data presented by Donko et al. \cite{Hartmann}.
    \label{fig:9}}
\end{figure}

\begin{figure}
  \centering
    \includegraphics[width=0.5\textwidth]{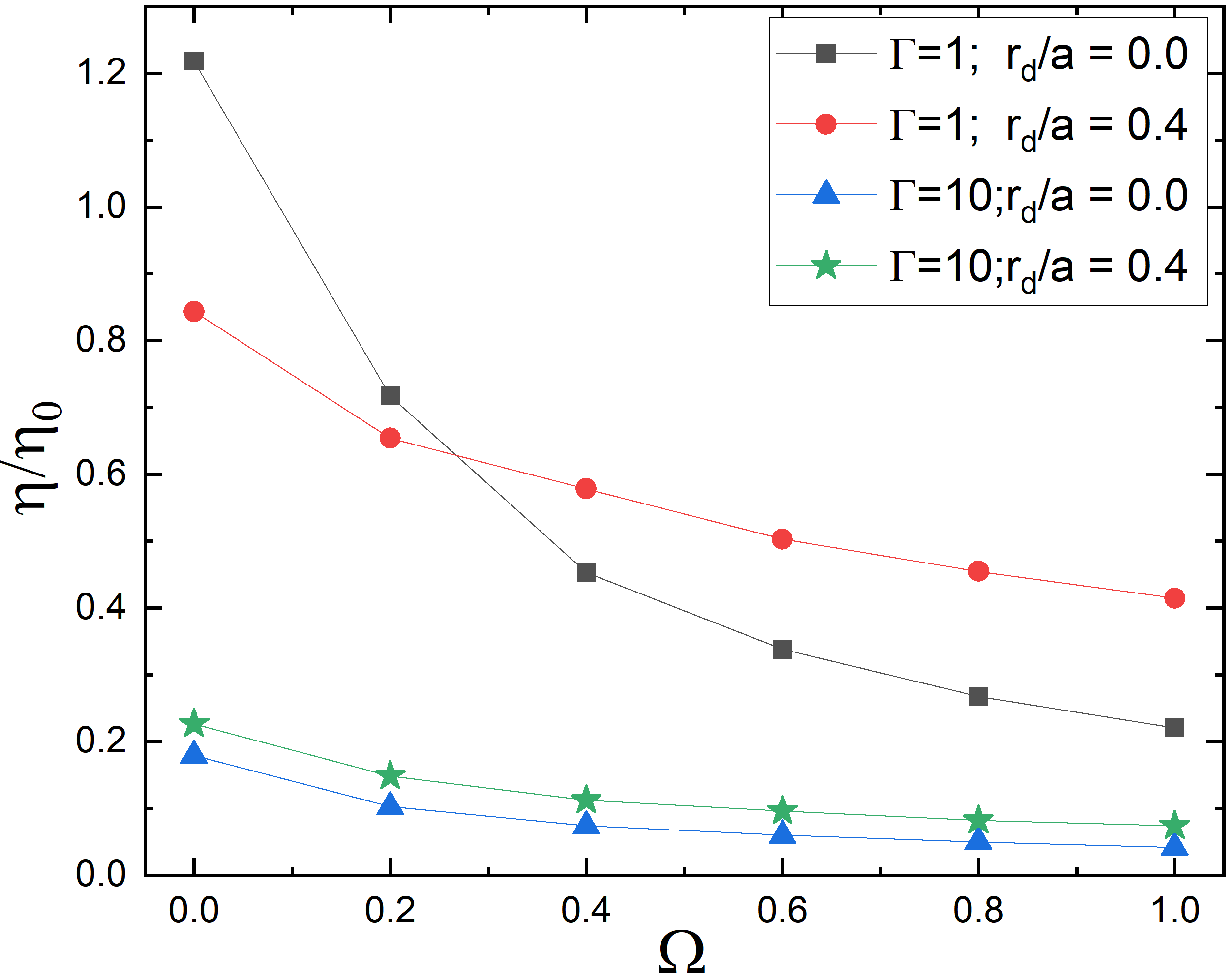}
        \caption{The dependence of the shear viscosity on the strength of the external magnetic field defined by $\Omega$ for $\Gamma = 1$, and $\Gamma = 10$. 
    \label{fig:15}}
\end{figure}

In Fig. \ref{fig:9}, we show the dependence of the shear viscosity on the coupling parameter without an external magnetic field. The results are computed using equilibrium and non-equilibrium MD methods without. From Fig. \ref{fig:9}, we see that the NEMD results are in good agreement with the EMD data. Moreover, the computed shear viscosity values are in close agreement with the results of Ref.\cite{Hartmann}. The general trend is that at $\Gamma\lesssim 40$, the increase in the coupling parameter leads to a decrease in the shear viscosity. 
From Fig. \ref{fig:9}, we see that the effect of taking into account the finite size of dust particles depends on the coupling parameter. At $\Gamma<2$, the shear viscosity of dust particles with $r_d=0.4a$ is reduced compared to the shear viscosity of point charges, i.e., the resistance to flow is decreased. This is correlated with the observation from the left panel of Fig. \ref{fig:7}, that the slope $\frac{\partial v_x}{\partial y}$ is larger for the system with $r_d=0.4a$ compared to the system of point charges (i.e., $r_d=0$). 
The smaller value of the shear viscosity is due to the inclusion of the effect of a finite size of dust particles. This can be interpreted as a result of a smaller number of close collisions that can transfer momentum in the most efficient way in this gas-like state. The behavior of the shear viscosity at small coupling parameters is known to be due to the kinetic term representing a correlation between momentums of particles at different times (the first term in the brackets on the r.h.s. of Eq. (\ref{eq:Stress tensor element}).  The increase in the probability of finding two particles at close distances (described by the RDF) is responsible for the increase in the shear viscosity with the decrease in $\Gamma$ at $\Gamma \ll 10$ \cite{Donko2}. Therefore, a decrease in the probability of finding two particles at a close distance due to finite-size effects also reduces the contribution of the kinetic term. 

In contrast, at $2<\Gamma\lesssim 30$, the shear viscosity of dust particles with $r_d=0.4a$ is larger than the shear viscosity of point charges. This is also correlated with the behavior of the slope  $\frac{\partial v_x}{\partial y}$  in the middle panel of Fig. \ref{fig:7}, from which we see that the model of point charges at $\Gamma=10$ has a larger magnitude of the derivative $\frac{\partial v_x}{\partial y}$  compared to that of the system of finite-sized charged particles. We interpret this as the result of the larger efficiency of the momentum transfer in the collision of particles due to taking into account that dust particles have a finite size. Indeed, at these parameters, the coupling is strong enough to create a short-range order in the system, meaning that a significant contribution to a momentum transfer is from collisions involving more than two particles (e.g., see Refs. \cite{Baalrud_pop_2014, Baalrud_prl_2013}  for the discussion of the connection between the RDF and transport properties).
At larger $\Gamma$ values, two considered systems have identical shear viscosity since the Coulomb repulsion is strong enough to keep particles at distances larger than $2r_d$ from each other.

From the presented results without an external magnetic field, we see that at considered parameters, taking into account the finite size of dust particles is important at $\Gamma\lesssim10$. Now, we consider how this effect manifests itself in the presence of an external magnetic field. For that, we calculated the shear viscosity coefficient using NEMD for $\Gamma=10$ and $\Gamma=1$ with $r_d=0$ and $r_d=0.4a$ at the magnetization parameters $\Omega\leq 1$. The results are presented in Fig. \ref{fig:15}. In general, the increase in the strength of the external magnetic field (here characterized by the parameter $\Omega$) leads to a decrease in the shear viscosity.  From Fig. \ref{fig:15}, it is clear that a stronger correlation of particles with a finite size reduces the influence of the external magnetic field on the shear viscosity compared to the system of point charges. Indeed, the stronger correlation between particles at $r_d=0.4a$ compared to the case with $r_d=0$ (see the RDFs in Fig. \ref{fig:RDF_Gamma}) means stronger localization of particles in respective potential minima, which leads to weakening of the influence of the Larmor rotation on the collision between particles and, correspondingly, on the momentum flux.

\subsection{Thermal conductivity}  \label{sec5b}
\begin{figure}
    \includegraphics[width=0.5\textwidth]{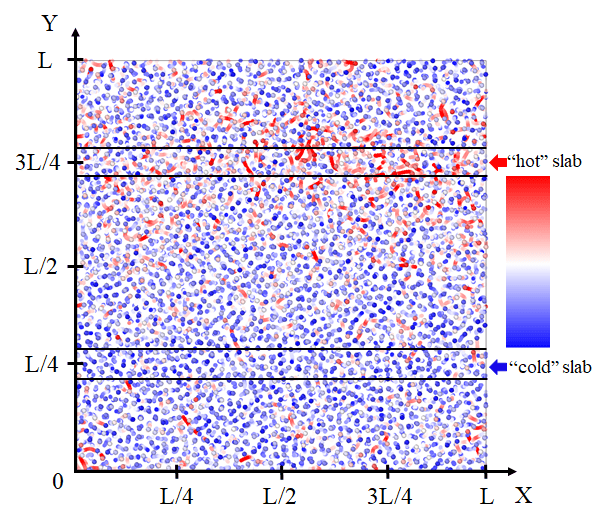}
        \caption{Color-coded particle trajectories at $\Gamma = 50$, $\kappa=2$, where the absolute value of the velocity increases from blue (relatively cold particles) to red (relatively hot particles).
    \label{fig:10}}
\end{figure}

A detailed description of the NEMD method for the calculation of thermal conductivity can be found in Ref. \cite{Müller}. According to the M\"uller-Plate method \cite{Müller}, we define two plates parallel to the x-axis where particles are heated and cooled to create a steady energy flux in the system. This is illustrated in Fig. \ref{fig:10}, where two slabs at heights of $y=(1/4)L$ and $y=(3/4)L$ are selected. To form a temperature gradient between these slabs, we find the particle with the largest velocity modulus in one slab and the particle with the smallest velocity modulus in the other slab and swap their momenta with a certain period. This generates a linear temperature distribution between heated and cooled slabs.
In Fig. \ref{fig:10}, the color-coded particle trajectories in a system with a coupling parameter $\Gamma = 50$ and a screening parameter $\kappa = 2$ are given. In the figure, colors are used to provide an indication of particle velocities: blue indicates low velocities and red indicates high velocities. 

\begin{figure*}[ht]
    \centering
    \begin{minipage}{.32\textwidth}
        \includegraphics[width=\linewidth]{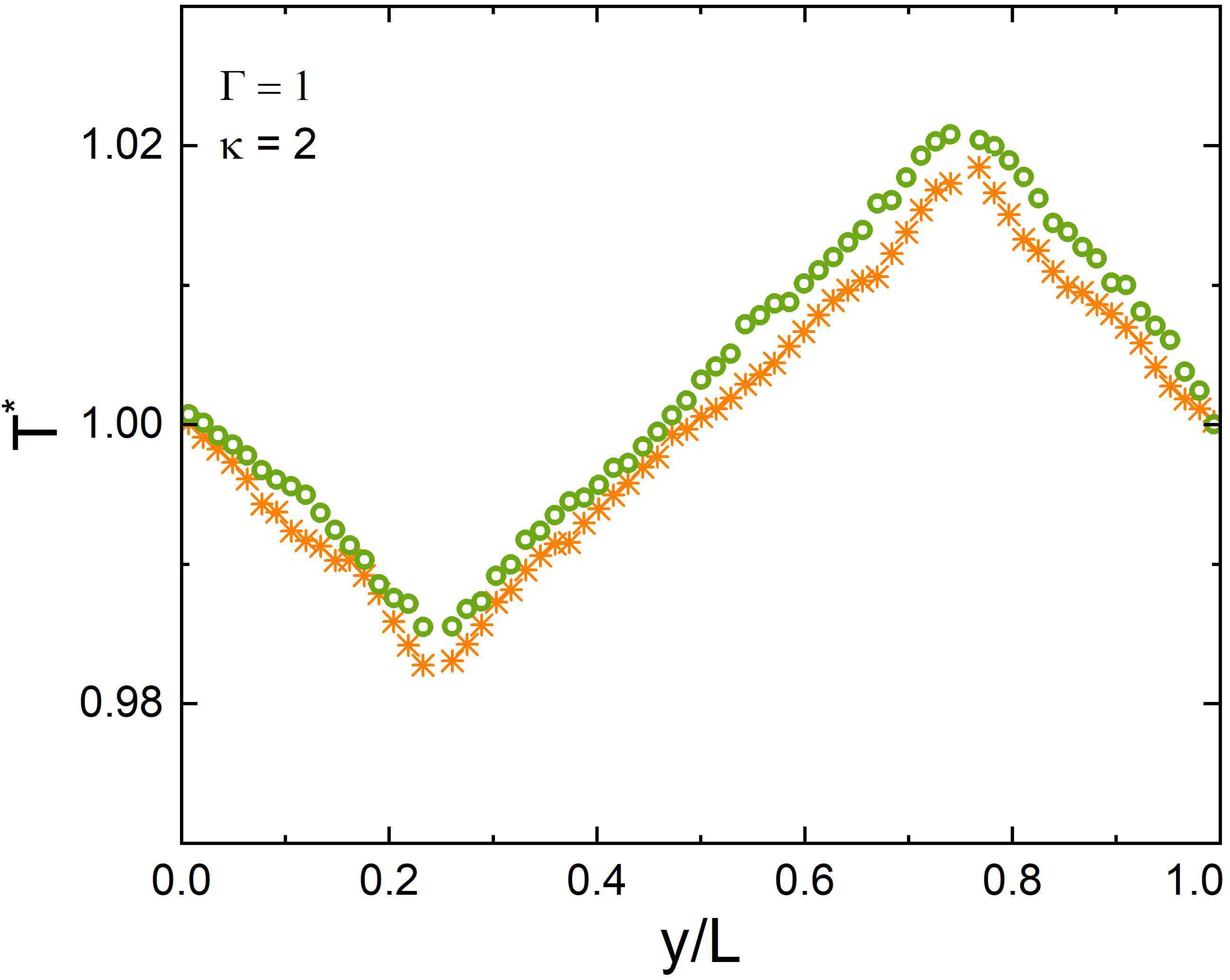}
        \centering
        \textbf{a)} 
    \end{minipage}%
    \hspace{1mm} 
    \begin{minipage}{.32\textwidth}
        \includegraphics[width=\linewidth]{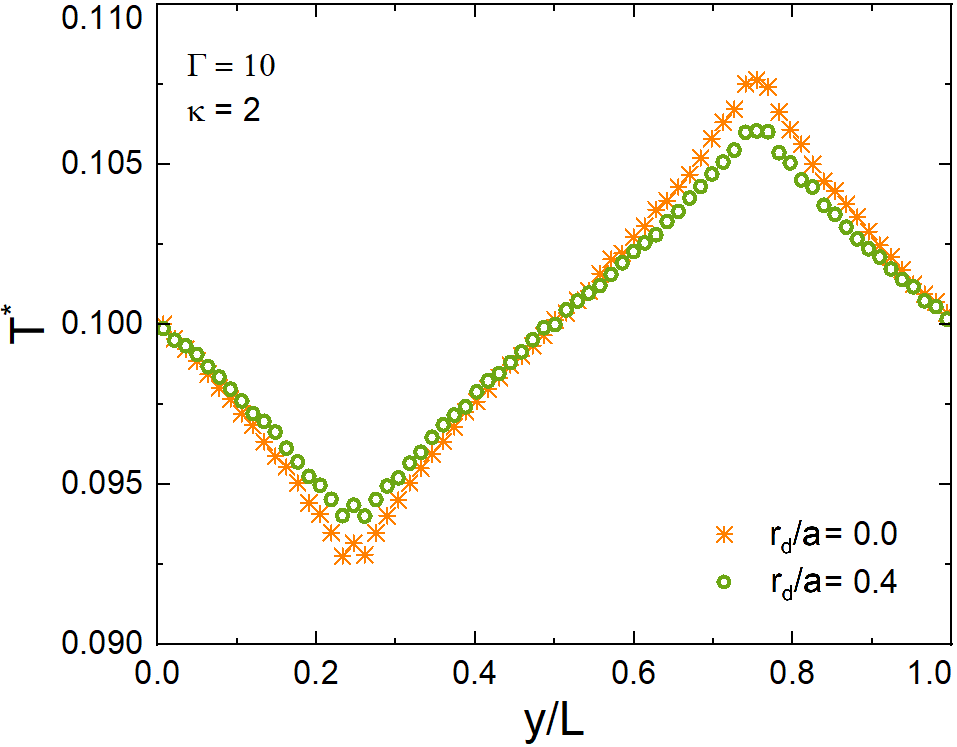}
        \centering
        \textbf{b)} 
    \end{minipage}%
     \hspace{1mm} 
       \begin{minipage}{.32\textwidth}
        \includegraphics[width=\linewidth]{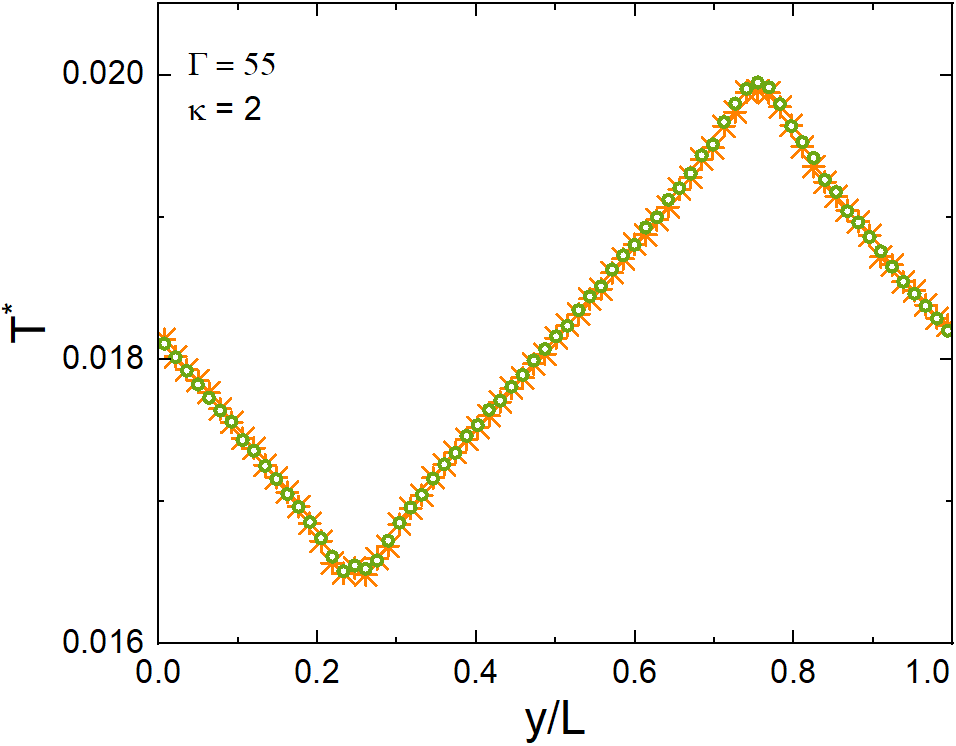}
        \centering
        \textbf{c)} 
    \end{minipage}%
    \caption{ Distribution of the temperature in the system between heated and cooled slabs for $\Gamma =1$, $10$, $55$, and $\kappa = 2$ at different values of the particle radius $r_{d}/a$.
    The calculations are performed without an external magnetic field.
    }
    \label{fig:11}
\end{figure*}

\begin{figure}
    \includegraphics[width=0.5\textwidth]{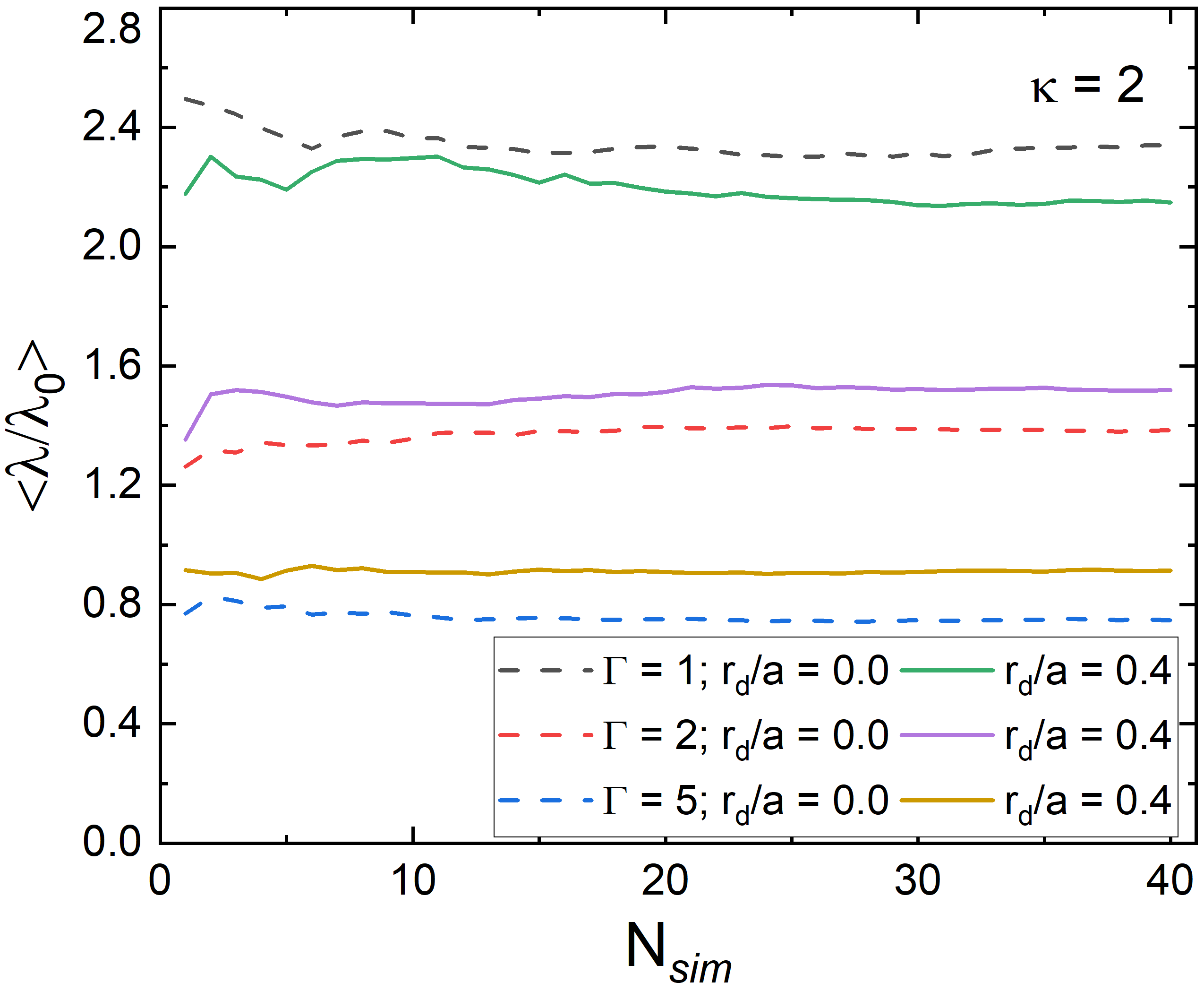}
        \caption{The dependence of the computed thermal conductivity on the number of the used values from independent simulations in the averaging. The results are for $r_d=0.0a$ and $r_d=0.4a$ at $\Gamma=1, 2, 5$. The calculations are performed without an external magnetic field.
    \label{fig:12}}
\end{figure}

Using information about the total value of the transferred thermal energy $\Delta E$  \cite{Müller} (i.e, the kinetic energy of chaotic motion of particles)  in a given time period $t$ and the temperature gradient $\frac{\partial T}{\partial y}$ from NEMD simulations, we computed the thermal conductivity using the following equation \cite{Müller}:
\begin{equation}\label{eq:lambda}
    \lambda = -\frac{\Delta E}{2Lt \frac{\partial T}{\partial y}}.
\end{equation}

The temperature gradient profiles for $\Gamma = 1$, $\Gamma=10$, and $\Gamma=55$ at $\kappa = 2$ and different values of particle radius are demonstrated in Fig.\ref{fig:11}.

To get reliable results for the thermal conductivity coefficient, we have performed 40 independent calculations for each data point and subsequently computed the average values. The convergence of the computed thermal conductivity with the increase in the number of independent runs are shown in  Fig.\ref{fig:12}. The result for the thermal conductivity are presented in units of $\lambda_0 = mn\omega_p a^2 k_B$.

\begin{figure}
    \includegraphics[width=0.5\textwidth]{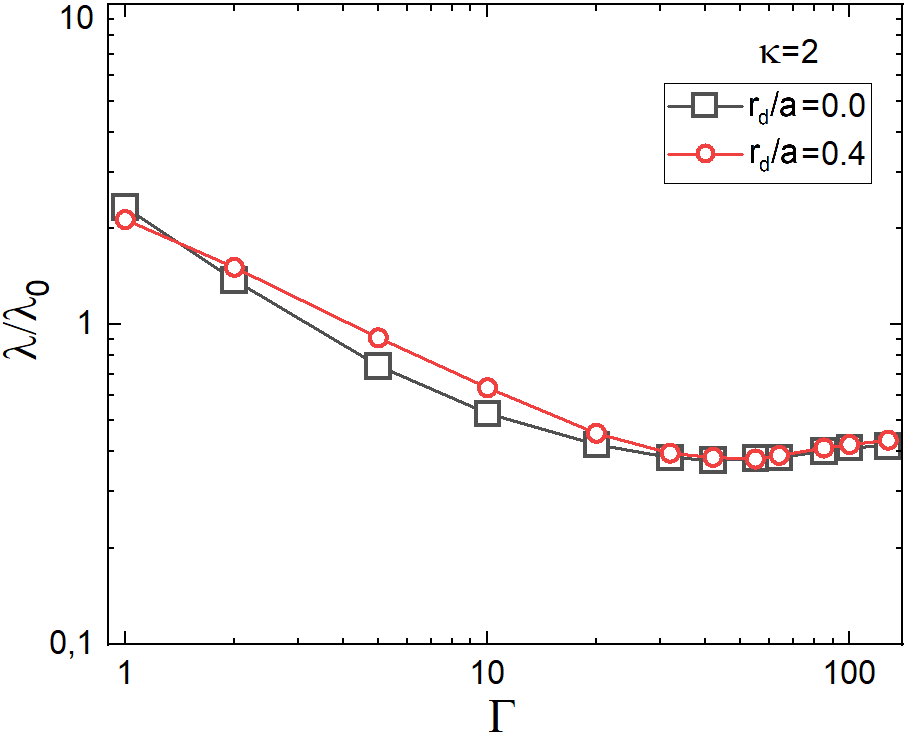}
        \caption{The thermal conductivity as a function of the coupling parameter $\Gamma$ for the 2D system of point charges ($r_d=0$) and the system of charged spheres with $r_d=0.4a$.
    \label{fig:13}}
\end{figure}

For the system of point charges  $r_{d}/a=0.0$ and for the system of particles with the radius $r_{d}/a=0.4$,
the calculated thermal conductivity values without an external magnetic field for different coupling parameters are shown in Fig.\ref{fig:13}. 
At $ \Gamma=1 $, the thermal conductivity for particles with $r_d=4 a$ is nearly identical to that of the system of point charges ( $r_d/a=0.0$).
At $2\leq \Gamma<30$, the thermal conductivity in the case $r_d=4 a$ is larger than that of the system of point charges. At larger coupling parameters $\Gamma>30$, two systems with $r_d=4 a$  and $r_d=0$ have equivalent thermal conductivity coefficients. The latter is due to the fact that the Coulomb repulsion is able to keep particles at large distances from each other, and the finite size of dust particles does not play any role in the transmission of energy across the system. On the other hand, at $2\leq \Gamma<30$, the finite size of particles with $r_d=0.4 a$ results in a stronger correlation of particles compared to the case with $r_d=0$, which leads to a more efficient thermal energy transmission. This follows from the middle panel of Fig.\ref{fig:11}, where one can see that the magnitude of $\frac{\partial T}{\partial y}$ is smaller for the system with $r_d=0.4a$ compared to the system with $r_d=0$. From Eq. (\ref{eq:lambda}), it is evident that thermal conductivity increases with the decrease in the magnitude of $\frac{\partial T}{\partial y}$ (for given values of $\Delta E$, $L$, and $t$). 

\begin{figure}
  \centering
    \includegraphics[width=0.5\textwidth]{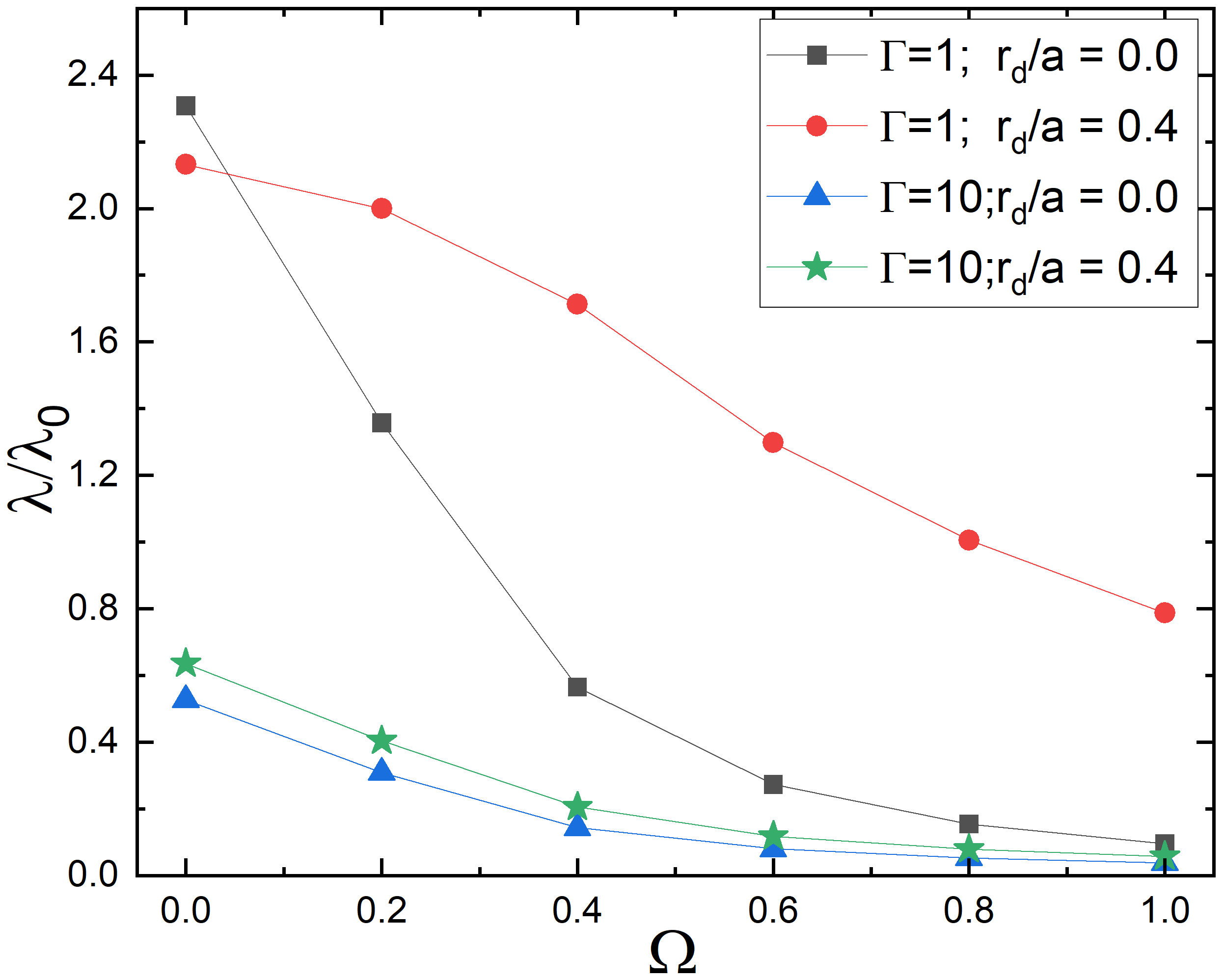}
        \caption{The dependence of the thermal conductivity on the strength of the external magnetic field defined by $\Omega$ for $\Gamma = 1$, and $\Gamma = 10$.
    \label{fig:14}}
\end{figure}

Now, let us analyze the change in thermal conductivity due to an external uniform magnetic field. In Fig.\ref{fig:14}, we show the results for the system of point charges and for the system of finite-sized charged particles with $r_d=0.4a$ at $\Gamma=1$, $\Gamma=10$ and the magnetization parameters $\Omega\leq 1$. Due to taking into account the finite size of dust particles, the effect of the magnetic field on the thermal conductivity is weakened compared to the case with $r_d=0$. 
This can be interpreted as a reduced impact of the particles' Larmor rotations on thermal conductivity due to the increased localization of particles when $r_d=0.4$, as opposed to the scenario with point charges.
The results indicate that in the regime of weak coupling, the finite size of particles leads to a different behavior of thermal conductivity compared to the model of point charges.

\section{Conclusions}\label{sec1}

We presented an approach to construct a force field that allows us to take into account a finite size of particles in MD simulations. In a general form Eq.(\ref{eq:1}), Eq. (\ref{eq:3}) and Eq. (\ref{eq:condition})  can be used in combination of other types of short-range repulsion force $\vec F_{\rm SR}(\vec r)$ and electrostatic force $\vec F_{E}(\vec r)$. For example, one can use an electrostatic interaction with non-linear screening \cite{Tsytovich2013, PhysRevLett.97.258302} or potentials taking into account higher-order terms in multipole expansion  \cite{PhysRevE.93.053204, boudec2024cartesiansphericalmultipoleexpansions}, which might be important for particles of non-spherical size \cite{7302071} or medium anisotropy \cite{boudec2024cartesiansphericalmultipoleexpansions}. Therefore, the presented approach has a wide range of utility for complex plasmas. The scheme presented can be used to investigate the effects of finite particle size on phenomena such as freezing and melting \cite{PhysRevB.60.14665, PhysRevE.86.051111, PhysRevE.102.033205}, diffusion \cite{PhysRevLett.103.195001}, and collective oscillation modes \cite{Bonitz}.

We demonstrated the proposed approach using the example of the finite-sized charged particles interacting via the Yukawa potential with screening parameter $\kappa=2$.  The generated pair interaction force between particles has been employed to compute the shear viscosity and thermal conductivity of the 2D system with and without an external magnetic field using non-equilibrium MD simulations. First, the results of the calculations show that the finite size of dust particles is important to be taken into account in the case of weak coupling $\Gamma\lesssim 10$. Second, the simulations show that the system of finite-sized charged particles can have larger or smaller shear viscosity depending on the value of the coupling parameter, which results from the interplay between kinetic (thermal) motion and correlation effects.  
Third, the fact that particles cannot overlap results in weaker sensitivity of the shear viscosity and thermal conductivity to the external magnetic field.  
These results are relevant to experiments where the effective magnetization of dust particles is realized by using the rotating frame approach \cite{PhysRevE.99.013203}.
The data shown in the paper are also provided in the tables in the appendix.

\section*{Acknowledgments}
This research is funded by the Science Committee of the Ministry of Education and Science of the republic of Kazakhstan (Grant AP19577935).

\section*{Data availability}
The values of the computed shear viscosity and thermal conductivity are provided in the appendix. The corresponding author can provide other data used and/or analyzed during the current study upon reasonable request.

\appendix

\section*{Appendix}
The values of the shear viscosity and the thermal conductivity presented in Fig.\ref{fig:9}, \ref{fig:15},\ref{fig:13} and \ref{fig:16} are provided in Tables \ref{tabl:tablichka1}-\ref{tabl:tablichka4}. The data for the momentum $j_x(p_x)$ and energy $j(\Delta E)$  fluxes corresponding to the parameters of the results presented in Fig. \ref{fig:7} and Fig. \ref{fig:11} are given in Table \ref{tabl:tablichka5} and Table \ref{tabl:tablichka6}, correspondingly.  

\begin{table}[h]
\caption{The shear viscosity values at different coupling parameters $\Gamma$ at $\kappa=2$, obtained by the NEMD and EMD methods as described in Sec. \ref{sec5a}.}
\label{tabl:tablichka1}
\begin{tabularx}{0.45\textwidth}{ | >{\centering\arraybackslash}X | >{\centering\arraybackslash}X | >
{\centering\arraybackslash}X | >
{\centering\arraybackslash}X | >{\centering\arraybackslash}X |}
\hline
$\Gamma$&$r_d=0.0$ &$r_d=0.0$&$r_d=0.4 a$&$r_d=0.4 a$\\ \hline
&NEMD&EMD&NEMD&EMD\\ \hline
1 & 1.218 & 1.236 & 0.843 & 0.769  \\
2 & 0.605 & 0.580 & 0.578 & 0.527 \\ 
3 & 0.424 & 0.396 & 0.466 & 0.426\\
5 & 0.282 & 0.281 & 0.348 & 0.330\\
10 & 0.179 & 0.186 & 0.227 & 0.228\\
20 & 0.127 & 0.121 & 0.145 & 0.134\\
32 & 0.109 & 0.107 & 0.114 & 0.097\\
42 & 0.106 & 0.095 & 0.108 & 0.099\\
55 & 0.106 & 0.109 & 0.105 & 0.09\\
64 & 0.107 & 0.107 & 0.107 & 0.107\\
85 & 0.113 & 0.108 & 0.113 &  0.108\\
100 & 0.120 & 0.118 & 0.120 & 0.118\\
128 & 0.135 & 0.141 & 0.135 & 0.141\\
\hline
\end{tabularx}
\end{table}
\begin{table}[h]
\caption{The shear viscosity values at different
strength of the external magnetic field defined by $\Omega$, for $\Gamma = 1$ and $\Gamma = 10$, obtained by the EMD method as described in Sec. \ref{sec5a}.}
\label{tabl:tablichka2}
\begin{tabularx}{0.45\textwidth}{ | >{\centering\arraybackslash}X | >{\centering\arraybackslash}X | >
{\centering\arraybackslash}X | >
{\centering\arraybackslash}X | >{\centering\arraybackslash}X |}
\hline
$\Omega$&$\Gamma = 1$ &$\Gamma = 1$ &$\Gamma = 10$ &$\Gamma = 10$ \\ \hline
&$r_d=0.0$  &$r_d=0.4 a$&$r_d=0.0$  &$r_d=0.4 a$\\ \hline
0 & 1.218 & 0.843 & 0.179 & 0.227  \\
0.2 & 0.718 & 0.654 & 0.103 & 0.148 \\ 
0.4 & 0.453 & 0.578 & 0.074 & 0.112\\
0.6 & 0.338 & 0.503 & 0.06 & 0.096\\
0.8 & 0.268 & 0.455 & 0.05 & 0.082\\
1.0 & 0.221 & 0.415 & 0.042 & 0.074\\
\hline
\end{tabularx}
\end{table}
\begin{table}[h]
\caption{The thermal conductivity values at different coupling parameters $\Gamma$ at  $\kappa=2$, obtained by the NEMD method as described in Sec. \ref{sec5b}.}
\label{tabl:tablichka3}
\begin{tabularx}{0.45\textwidth}{ | >{\centering\arraybackslash}X | >{\centering\arraybackslash}X | >
{\centering\arraybackslash}X | >
{\centering\arraybackslash}X | >{\centering\arraybackslash}X |}
\hline
$\Gamma$&$r_d=0.0$&$r_d=0.4 a$\\ \hline
1 & 2.326 & 2.149   \\
2 & 1.379 & 1.519  \\ 
5 & 0.745 & 0.902 \\
10 & 0.529 & 0.632 \\
20 & 0.421 & 0.452 \\
32 & 0.389 & 0.399 \\
42 & 0.381 & 0.384 \\
55 & 0.384 & 0.379 \\
64 & 0.384 & 0.392 \\
85 & 0.404 & 0.410 \\
100 & 0.416 & 0.422 \\
128 & 0.413 & 0.434 \\
\hline
\end{tabularx}
\end{table}

\begin{table}[h]
\caption{The thermal conductivity values at different
strength of the external magnetic field defined by $\Omega$, for $\Gamma = 1$ and $\Gamma = 10$, obtained by the NEMD method as described in Sec. \ref{sec5b}.}
\label{tabl:tablichka4}
\begin{tabularx}{0.45\textwidth}{ | >{\centering\arraybackslash}X | >{\centering\arraybackslash}X | >
{\centering\arraybackslash}X | >
{\centering\arraybackslash}X | >{\centering\arraybackslash}X |}
\hline
$\Omega$&$\Gamma = 1$ &$\Gamma = 1$ &$\Gamma = 10$ &$\Gamma = 10$ \\ \hline
&$r_d=0.0$  &$r_d=0.4 a$&$r_d=0.0$  &$r_d=0.4 a$\\ \hline
0 & 2.308 & 2.132 & 0.527 & 0.636  \\
0.2 & 1.358 & 2.047 & 0.309 & 0.404 \\ 
0.4 & 0.565 & 1.713 & 0.144 & 0.205\\
0.6 & 0.273 & 1.298 & 0.079 & 0.118\\
0.8 & 0.155 & 1.006 & 0.053 & 0.079\\
1.0 & 0.095 & 0.787 & 0.038 & 0.057\\
\hline
\end{tabularx}
\end{table}
{
The effect of different short-range repulsion models on the shear viscosity is demonstrated in Fig.\ref{fig:16}. As can be seen from the figure Fig.\ref{fig:16}, the values of shear viscosity for all considered coupling parameter values are in agreement with each other when different short-range repulsion models are used.


As a numerical demonstration of the fact that the external homogenous magnetic field does not affect the RDF, in Fig. \ref{fig:18}, we show the radial distribution function (RDF) of particles in the presence of the external magnetic field with the magnetization parameter $\Omega=1$ in comparison with the RDF  obtained without an external magnetic field.  As can be seen from Fig. \ref{fig:18}, the external magnetic field does not affect the RDF values.

\begin{figure}
  \centering
    \includegraphics[width=0.5\textwidth]{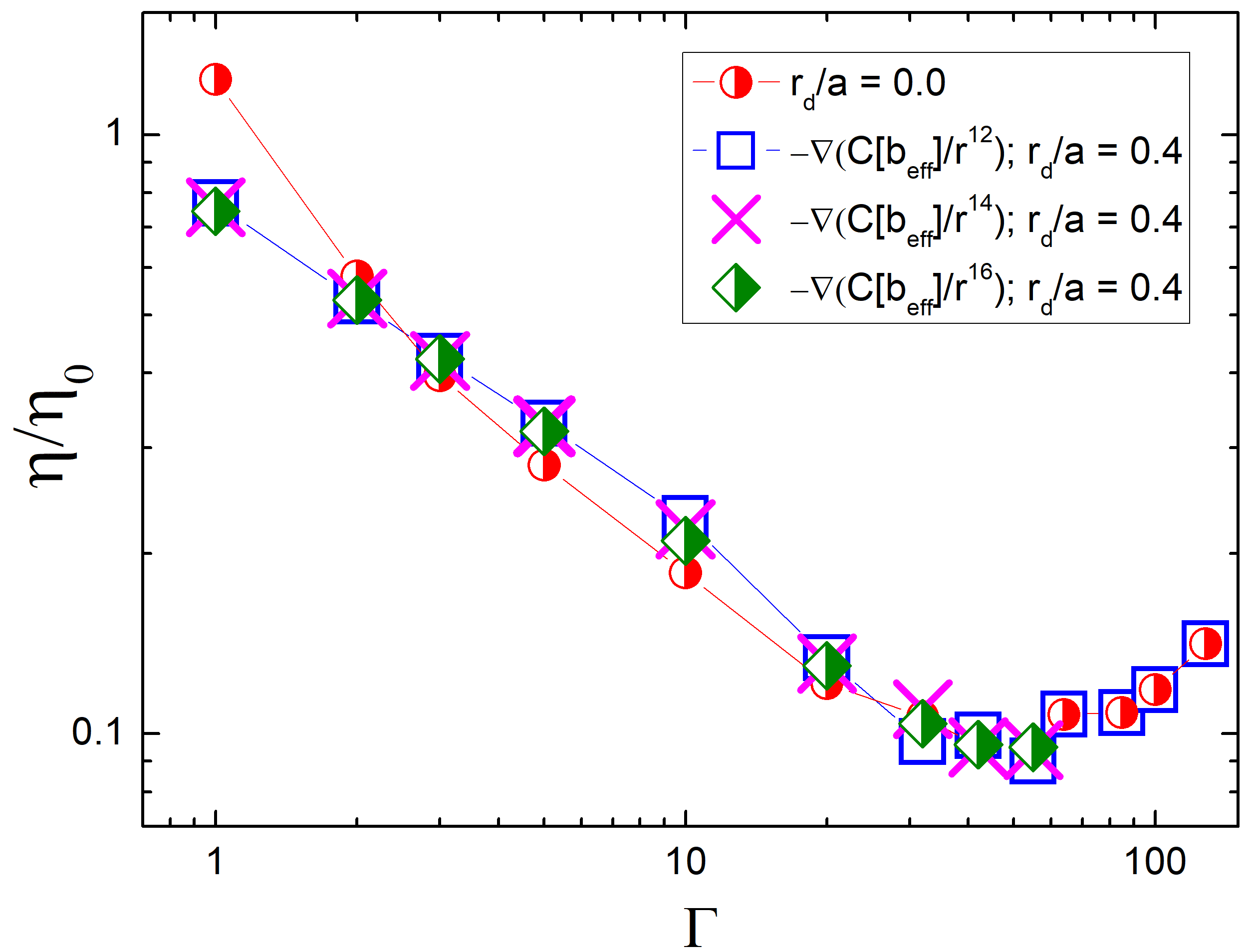}
        \caption{ The dependence of shear viscosity on the coupling parameter (b) for the different short-range repulsion models. The results are presented for different values of the particle radius $r_{d}/a=0.0$ and $r_{d}/a=0.4$.}
    \label{fig:16}
\end{figure}

\begin{figure}
  \centering
    \includegraphics[width=0.5\textwidth]{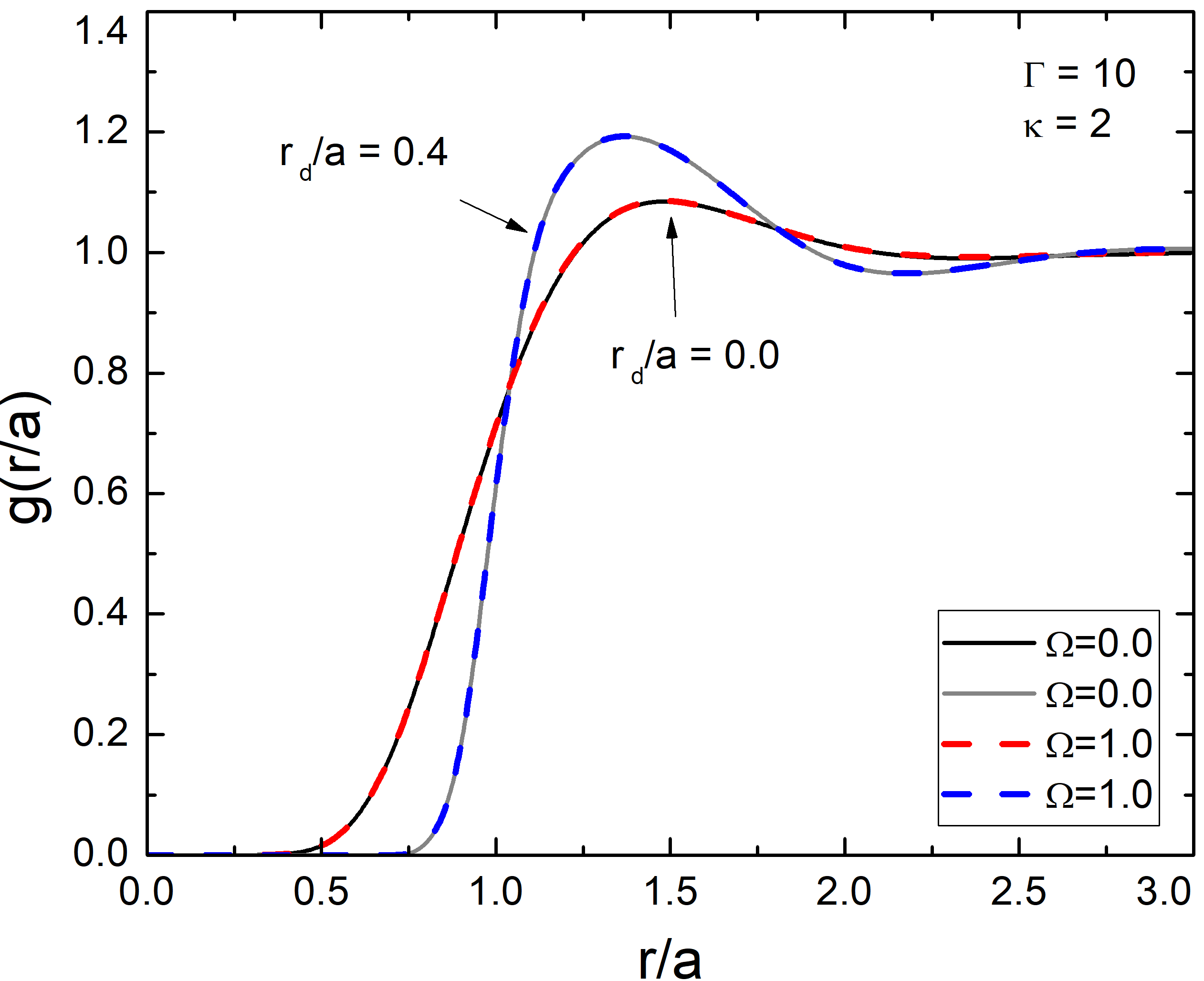}
        \caption{The radial distribution function (RDF) for the strength of the external magnetic field defined by $\Omega=1.0$ and without external magnetic field $\Omega=0.0$  for $\Gamma = 10$ and $\kappa = 2$.}
    \label{fig:18}
\end{figure}
}

\begin{table}[h]
\caption{The momentum flux for different $\Gamma$ and for different values of the particle radius $r_{d}/a=0.0$ and $r_{d}/a=0.4$, obtained by the NEMD method as described in Sec. \ref{sec5a}. The results were obtained after averaging over $N_{sim}$ independent simulations.}
\label{tabl:tablichka5}
\begin{tabularx}{0.7\textwidth}{ | >{\centering\arraybackslash}X | >{\centering\arraybackslash}X | >
{\centering\arraybackslash}X | >
{\centering\arraybackslash}X | >
{\centering\arraybackslash}X | >{\centering\arraybackslash}X | >
{\centering\arraybackslash}X |}
\hline
$\Gamma$&$t_{sim} \omega_p$&$N_{sim}$&$\Delta p$  &$\Delta p$&$j_x(p_x)$  &$j_x(p_x)$ \\ \hline
& & & $r_d=0.0$  &$r_d=0.4 a$&$r_d=0.0$  &$r_d=0.4 a$\\ \hline
1&  60000& 100 &14482.3 & 14355.8 & 0.121 & 0.12  \\
10 & 10000& 50 &552.21 & 595.21 & 0.028 & 0.03 \\ 
55 &10000& 50 &191.33 & 190.68 & 0.0096 & 0.0095\\
\hline
\end{tabularx}
\end{table}
\begin{table}[h]
\caption{The energy flux for different $\Gamma$ and for different values of the particle radius $r_{d}/a=0.0$ and $r_{d}/a=0.4$, obtained by the NEMD method as described in Sec. \ref{sec5b}. The results were obtained after averaging over $N_{sim}$ independent simulations.}
\label{tabl:tablichka6}
\begin{tabularx}{0.7\textwidth}{ | >{\centering\arraybackslash}X | >
{\centering\arraybackslash}X | >{\centering\arraybackslash}X | >
{\centering\arraybackslash}X | >
{\centering\arraybackslash}X | >{\centering\arraybackslash}X | >
{\centering\arraybackslash}X |}
\hline
$\Gamma$&$t_{sim} \omega_p$&$N_{sim}$&$\Delta E$  &$\Delta E$&$j(\Delta E)$  &$j(\Delta E)$ \\ \hline
& & & $r_d=0.0$  &$r_d=0.4 a$&$r_d=0.0$  &$r_d=0.4 a$\\ \hline
1&  60000& 40 & 6032.43 & 6072.33 & 0.05 & 0.05  \\
10 & 30000& 40 & 284.99 & 287.33 & 0.0047 & 0.0048 \\ 
55 &30000& 40 & 49.79 & 49.89 & 0.00083 & 0.000832\\
\hline
\end{tabularx}
\end{table}

 \bibliography{ref}

\begin{thebibliography}{62}%
\makeatletter
\providecommand \@ifxundefined [1]{%
 \@ifx{#1\undefined}
}%
\providecommand \@ifnum [1]{%
 \ifnum #1\expandafter \@firstoftwo
 \else \expandafter \@secondoftwo
 \fi
}%
\providecommand \@ifx [1]{%
 \ifx #1\expandafter \@firstoftwo
 \else \expandafter \@secondoftwo
 \fi
}%
\providecommand \natexlab [1]{#1}%
\providecommand \enquote  [1]{``#1''}%
\providecommand \bibnamefont  [1]{#1}%
\providecommand \bibfnamefont [1]{#1}%
\providecommand \citenamefont [1]{#1}%
\providecommand \href@noop [0]{\@secondoftwo}%
\providecommand \href [0]{\begingroup \@sanitize@url \@href}%
\providecommand \@href[1]{\@@startlink{#1}\@@href}%
\providecommand \@@href[1]{\endgroup#1\@@endlink}%
\providecommand \@sanitize@url [0]{\catcode `\\12\catcode `\$12\catcode `\&12\catcode `\#12\catcode `\^12\catcode `\_12\catcode `\%12\relax}%
\providecommand \@@startlink[1]{}%
\providecommand \@@endlink[0]{}%
\providecommand \url  [0]{\begingroup\@sanitize@url \@url }%
\providecommand \@url [1]{\endgroup\@href {#1}{\urlprefix }}%
\providecommand \urlprefix  [0]{URL }%
\providecommand \Eprint [0]{\href }%
\providecommand \doibase [0]{http://dx.doi.org/}%
\providecommand \selectlanguage [0]{\@gobble}%
\providecommand \bibinfo  [0]{\@secondoftwo}%
\providecommand \bibfield  [0]{\@secondoftwo}%
\providecommand \translation [1]{[#1]}%
\providecommand \BibitemOpen [0]{}%
\providecommand \bibitemStop [0]{}%
\providecommand \bibitemNoStop [0]{.\EOS\space}%
\providecommand \EOS [0]{\spacefactor3000\relax}%
\providecommand \BibitemShut  [1]{\csname bibitem#1\endcsname}%
\let\auto@bib@innerbib\@empty
\bibitem [{\citenamefont {Wong}\ \emph {et~al.}(2018)\citenamefont {Wong}, \citenamefont {Goree}, \citenamefont {Haralson},\ and\ \citenamefont {Liu}}]{Wong2018}%
  \BibitemOpen
  \bibfield  {author} {\bibinfo {author} {\bibfnamefont {C.-S.}\ \bibnamefont {Wong}}, \bibinfo {author} {\bibfnamefont {J.}~\bibnamefont {Goree}}, \bibinfo {author} {\bibfnamefont {Z.}~\bibnamefont {Haralson}}, \ and\ \bibinfo {author} {\bibfnamefont {B.}~\bibnamefont {Liu}},\ }\href {\doibase 10.1038/nphys4253} {\bibfield  {journal} {\bibinfo  {journal} {Nature Physics}\ }\textbf {\bibinfo {volume} {14}},\ \bibinfo {pages} {21} (\bibinfo {year} {2018})}\BibitemShut {NoStop}%
\bibitem [{\citenamefont {Petersen}\ \emph {et~al.}(2022)\citenamefont {Petersen}, \citenamefont {Asnaz}, \citenamefont {Tadsen},\ and\ \citenamefont {Greiner}}]{Petersen2022}%
  \BibitemOpen
  \bibfield  {author} {\bibinfo {author} {\bibfnamefont {A.}~\bibnamefont {Petersen}}, \bibinfo {author} {\bibfnamefont {O.~H.}\ \bibnamefont {Asnaz}}, \bibinfo {author} {\bibfnamefont {B.}~\bibnamefont {Tadsen}}, \ and\ \bibinfo {author} {\bibfnamefont {F.}~\bibnamefont {Greiner}},\ }\href {\doibase 10.1038/s42005-022-01060-5} {\bibfield  {journal} {\bibinfo  {journal} {Communications Physics}\ }\textbf {\bibinfo {volume} {5}},\ \bibinfo {pages} {308} (\bibinfo {year} {2022})}\BibitemShut {NoStop}%
\bibitem [{\citenamefont {Fortov}\ \emph {et~al.}(2005)\citenamefont {Fortov}, \citenamefont {Ivlev}, \citenamefont {Khrapak}, \citenamefont {Khrapak},\ and\ \citenamefont {Morfill}}]{FORTOV20051}%
  \BibitemOpen
  \bibfield  {author} {\bibinfo {author} {\bibfnamefont {V.}~\bibnamefont {Fortov}}, \bibinfo {author} {\bibfnamefont {A.}~\bibnamefont {Ivlev}}, \bibinfo {author} {\bibfnamefont {S.}~\bibnamefont {Khrapak}}, \bibinfo {author} {\bibfnamefont {A.}~\bibnamefont {Khrapak}}, \ and\ \bibinfo {author} {\bibfnamefont {G.}~\bibnamefont {Morfill}},\ }\href {\doibase https://doi.org/10.1016/j.physrep.2005.08.007} {\bibfield  {journal} {\bibinfo  {journal} {Physics Reports}\ }\textbf {\bibinfo {volume} {421}},\ \bibinfo {pages} {1} (\bibinfo {year} {2005})}\BibitemShut {NoStop}%
\bibitem [{\citenamefont {Likos}(2001)}]{LIKOS2001267}%
  \BibitemOpen
  \bibfield  {author} {\bibinfo {author} {\bibfnamefont {C.~N.}\ \bibnamefont {Likos}},\ }\href {\doibase https://doi.org/10.1016/S0370-1573(00)00141-1} {\bibfield  {journal} {\bibinfo  {journal} {Physics Reports}\ }\textbf {\bibinfo {volume} {348}},\ \bibinfo {pages} {267} (\bibinfo {year} {2001})}\BibitemShut {NoStop}%
\bibitem [{\citenamefont {Stradner}\ \emph {et~al.}(2004)\citenamefont {Stradner}, \citenamefont {Sedgwick}, \citenamefont {Cardinaux}, \citenamefont {Poon}, \citenamefont {Egelhaaf},\ and\ \citenamefont {Schurtenberger}}]{Stradner2004}%
  \BibitemOpen
  \bibfield  {author} {\bibinfo {author} {\bibfnamefont {A.}~\bibnamefont {Stradner}}, \bibinfo {author} {\bibfnamefont {H.}~\bibnamefont {Sedgwick}}, \bibinfo {author} {\bibfnamefont {F.}~\bibnamefont {Cardinaux}}, \bibinfo {author} {\bibfnamefont {W.~C.~K.}\ \bibnamefont {Poon}}, \bibinfo {author} {\bibfnamefont {S.~U.}\ \bibnamefont {Egelhaaf}}, \ and\ \bibinfo {author} {\bibfnamefont {P.}~\bibnamefont {Schurtenberger}},\ }\href {\doibase 10.1038/nature03109} {\bibfield  {journal} {\bibinfo  {journal} {Nature}\ }\textbf {\bibinfo {volume} {432}},\ \bibinfo {pages} {492} (\bibinfo {year} {2004})}\BibitemShut {NoStop}%
\bibitem [{\citenamefont {Hartmann}\ \emph {et~al.}(2005)\citenamefont {Hartmann}, \citenamefont {Kalman}, \citenamefont {Donk\'o},\ and\ \citenamefont {Kutasi}}]{Hartmann2005}%
  \BibitemOpen
  \bibfield  {author} {\bibinfo {author} {\bibfnamefont {P.}~\bibnamefont {Hartmann}}, \bibinfo {author} {\bibfnamefont {G.~J.}\ \bibnamefont {Kalman}}, \bibinfo {author} {\bibfnamefont {Z.}~\bibnamefont {Donk\'o}}, \ and\ \bibinfo {author} {\bibfnamefont {K.}~\bibnamefont {Kutasi}},\ }\href {\doibase 10.1103/PhysRevE.72.026409} {\bibfield  {journal} {\bibinfo  {journal} {Phys. Rev. E}\ }\textbf {\bibinfo {volume} {72}},\ \bibinfo {pages} {026409} (\bibinfo {year} {2005})}\BibitemShut {NoStop}%
\bibitem [{\citenamefont {Vaulina}\ and\ \citenamefont {Koss}(2015)}]{Vaulina2015}%
  \BibitemOpen
  \bibfield  {author} {\bibinfo {author} {\bibfnamefont {O.~S.}\ \bibnamefont {Vaulina}}\ and\ \bibinfo {author} {\bibfnamefont {X.~G.}\ \bibnamefont {Koss}},\ }\href {\doibase 10.1103/PhysRevE.92.042155} {\bibfield  {journal} {\bibinfo  {journal} {Phys. Rev. E}\ }\textbf {\bibinfo {volume} {92}},\ \bibinfo {pages} {042155} (\bibinfo {year} {2015})}\BibitemShut {NoStop}%
\bibitem [{\citenamefont {Hamaguchi}\ \emph {et~al.}(1997)\citenamefont {Hamaguchi}, \citenamefont {Farouki},\ and\ \citenamefont {Dubin}}]{Hamaguchi1997}%
  \BibitemOpen
  \bibfield  {author} {\bibinfo {author} {\bibfnamefont {S.}~\bibnamefont {Hamaguchi}}, \bibinfo {author} {\bibfnamefont {R.~T.}\ \bibnamefont {Farouki}}, \ and\ \bibinfo {author} {\bibfnamefont {D.~H.~E.}\ \bibnamefont {Dubin}},\ }\href {\doibase 10.1103/PhysRevE.56.4671} {\bibfield  {journal} {\bibinfo  {journal} {Phys. Rev. E}\ }\textbf {\bibinfo {volume} {56}},\ \bibinfo {pages} {4671} (\bibinfo {year} {1997})}\BibitemShut {NoStop}%
\bibitem [{\citenamefont {Khrustalyov}\ and\ \citenamefont {Vaulina}(2011)}]{Khrustalyov2011}%
  \BibitemOpen
  \bibfield  {author} {\bibinfo {author} {\bibfnamefont {Y.~V.}\ \bibnamefont {Khrustalyov}}\ and\ \bibinfo {author} {\bibfnamefont {O.~S.}\ \bibnamefont {Vaulina}},\ }\href {https://arxiv.org/abs/1112.0064} {\bibfield  {journal} {\bibinfo  {journal} {arXiv}\ } (\bibinfo {year} {2011})}\BibitemShut {NoStop}%
\bibitem [{\citenamefont {Liu}\ and\ \citenamefont {Goree}(2005)}]{Liu2005}%
  \BibitemOpen
  \bibfield  {author} {\bibinfo {author} {\bibfnamefont {B.}~\bibnamefont {Liu}}\ and\ \bibinfo {author} {\bibfnamefont {J.}~\bibnamefont {Goree}},\ }\href {\doibase 10.1103/PhysRevLett.94.185002} {\bibfield  {journal} {\bibinfo  {journal} {Phys. Rev. Lett.}\ }\textbf {\bibinfo {volume} {94}},\ \bibinfo {pages} {185002} (\bibinfo {year} {2005})}\BibitemShut {NoStop}%
\bibitem [{\citenamefont {Donk\'o}\ and\ \citenamefont {Hartmann}(2004)}]{PhysRevE.69.016405}%
  \BibitemOpen
  \bibfield  {author} {\bibinfo {author} {\bibfnamefont {Z.}~\bibnamefont {Donk\'o}}\ and\ \bibinfo {author} {\bibfnamefont {P.}~\bibnamefont {Hartmann}},\ }\href {\doibase 10.1103/PhysRevE.69.016405} {\bibfield  {journal} {\bibinfo  {journal} {Phys. Rev. E}\ }\textbf {\bibinfo {volume} {69}},\ \bibinfo {pages} {016405} (\bibinfo {year} {2004})}\BibitemShut {NoStop}%
\bibitem [{\citenamefont {Kodanova}\ \emph {et~al.}(2015)\citenamefont {Kodanova}, \citenamefont {Ramazanov}, \citenamefont {Bastykova},\ and\ \citenamefont {Moldabekov}}]{10.1063/1.4922908}%
  \BibitemOpen
  \bibfield  {author} {\bibinfo {author} {\bibfnamefont {S.~K.}\ \bibnamefont {Kodanova}}, \bibinfo {author} {\bibfnamefont {T.~S.}\ \bibnamefont {Ramazanov}}, \bibinfo {author} {\bibfnamefont {N.~K.}\ \bibnamefont {Bastykova}}, \ and\ \bibinfo {author} {\bibfnamefont {Z.~A.}\ \bibnamefont {Moldabekov}},\ }\href {\doibase 10.1063/1.4922908} {\bibfield  {journal} {\bibinfo  {journal} {Physics of Plasmas}\ }\textbf {\bibinfo {volume} {22}},\ \bibinfo {pages} {063703} (\bibinfo {year} {2015})},\ \Eprint {http://arxiv.org/abs/https://pubs.aip.org/aip/pop/article-pdf/doi/10.1063/1.4922908/15863243/063703\_1\_online.pdf} {https://pubs.aip.org/aip/pop/article-pdf/doi/10.1063/1.4922908/15863243/063703\_1\_online.pdf} \BibitemShut {NoStop}%
\bibitem [{\citenamefont {Djienbekov}\ \emph {et~al.}(2022{\natexlab{a}})\citenamefont {Djienbekov}, \citenamefont {Bastykova}, \citenamefont {Bekbussyn}, \citenamefont {Ramazanov},\ and\ \citenamefont {Kodanova}}]{PhysRevE.106.065203}%
  \BibitemOpen
  \bibfield  {author} {\bibinfo {author} {\bibfnamefont {N.~E.}\ \bibnamefont {Djienbekov}}, \bibinfo {author} {\bibfnamefont {N.~K.}\ \bibnamefont {Bastykova}}, \bibinfo {author} {\bibfnamefont {A.~M.}\ \bibnamefont {Bekbussyn}}, \bibinfo {author} {\bibfnamefont {T.~S.}\ \bibnamefont {Ramazanov}}, \ and\ \bibinfo {author} {\bibfnamefont {S.~K.}\ \bibnamefont {Kodanova}},\ }\href {\doibase 10.1103/PhysRevE.106.065203} {\bibfield  {journal} {\bibinfo  {journal} {Phys. Rev. E}\ }\textbf {\bibinfo {volume} {106}},\ \bibinfo {pages} {065203} (\bibinfo {year} {2022}{\natexlab{a}})}\BibitemShut {NoStop}%
\bibitem [{\citenamefont {Feng}\ \emph {et~al.}(2012)\citenamefont {Feng}, \citenamefont {Goree},\ and\ \citenamefont {Liu}}]{PhysRevLett.109.185002}%
  \BibitemOpen
  \bibfield  {author} {\bibinfo {author} {\bibfnamefont {Y.}~\bibnamefont {Feng}}, \bibinfo {author} {\bibfnamefont {J.}~\bibnamefont {Goree}}, \ and\ \bibinfo {author} {\bibfnamefont {B.}~\bibnamefont {Liu}},\ }\href {\doibase 10.1103/PhysRevLett.109.185002} {\bibfield  {journal} {\bibinfo  {journal} {Phys. Rev. Lett.}\ }\textbf {\bibinfo {volume} {109}},\ \bibinfo {pages} {185002} (\bibinfo {year} {2012})}\BibitemShut {NoStop}%
\bibitem [{\citenamefont {Moldabekov}\ \emph {et~al.}(2015)\citenamefont {Moldabekov}, \citenamefont {Schoof}, \citenamefont {Ludwig}, \citenamefont {Bonitz},\ and\ \citenamefont {Ramazanov}}]{10.1063/1.4932051}%
  \BibitemOpen
  \bibfield  {author} {\bibinfo {author} {\bibfnamefont {Z.}~\bibnamefont {Moldabekov}}, \bibinfo {author} {\bibfnamefont {T.}~\bibnamefont {Schoof}}, \bibinfo {author} {\bibfnamefont {P.}~\bibnamefont {Ludwig}}, \bibinfo {author} {\bibfnamefont {M.}~\bibnamefont {Bonitz}}, \ and\ \bibinfo {author} {\bibfnamefont {T.}~\bibnamefont {Ramazanov}},\ }\href {\doibase 10.1063/1.4932051} {\bibfield  {journal} {\bibinfo  {journal} {Physics of Plasmas}\ }\textbf {\bibinfo {volume} {22}},\ \bibinfo {pages} {102104} (\bibinfo {year} {2015})},\ \Eprint {http://arxiv.org/abs/https://pubs.aip.org/aip/pop/article-pdf/doi/10.1063/1.4932051/13527324/102104\_1\_online.pdf} {https://pubs.aip.org/aip/pop/article-pdf/doi/10.1063/1.4932051/13527324/102104\_1\_online.pdf} \BibitemShut {NoStop}%
\bibitem [{\citenamefont {Moldabekov}\ \emph {et~al.}(2017)\citenamefont {Moldabekov}, \citenamefont {Groth}, \citenamefont {Dornheim}, \citenamefont {Bonitz},\ and\ \citenamefont {Ramazanov}}]{Moldabekov_cpp_2017}%
  \BibitemOpen
  \bibfield  {author} {\bibinfo {author} {\bibfnamefont {Z.}~\bibnamefont {Moldabekov}}, \bibinfo {author} {\bibfnamefont {S.}~\bibnamefont {Groth}}, \bibinfo {author} {\bibfnamefont {T.}~\bibnamefont {Dornheim}}, \bibinfo {author} {\bibfnamefont {M.}~\bibnamefont {Bonitz}}, \ and\ \bibinfo {author} {\bibfnamefont {T.}~\bibnamefont {Ramazanov}},\ }\href {\doibase https://doi.org/10.1002/ctpp.201700109} {\bibfield  {journal} {\bibinfo  {journal} {Contributions to Plasma Physics}\ }\textbf {\bibinfo {volume} {57}},\ \bibinfo {pages} {532} (\bibinfo {year} {2017})},\ \Eprint {http://arxiv.org/abs/https://onlinelibrary.wiley.com/doi/pdf/10.1002/ctpp.201700109} {https://onlinelibrary.wiley.com/doi/pdf/10.1002/ctpp.201700109} \BibitemShut {NoStop}%
\bibitem [{\citenamefont {Moldabekov}\ \emph {et~al.}(2012)\citenamefont {Moldabekov}, \citenamefont {Ramazanov},\ and\ \citenamefont {Dzhumagulova}}]{Moldabekov_cpp_2012}%
  \BibitemOpen
  \bibfield  {author} {\bibinfo {author} {\bibfnamefont {Z.~A.}\ \bibnamefont {Moldabekov}}, \bibinfo {author} {\bibfnamefont {T.~S.}\ \bibnamefont {Ramazanov}}, \ and\ \bibinfo {author} {\bibfnamefont {K.~N.}\ \bibnamefont {Dzhumagulova}},\ }\href {\doibase https://doi.org/10.1002/ctpp.201100072} {\bibfield  {journal} {\bibinfo  {journal} {Contributions to Plasma Physics}\ }\textbf {\bibinfo {volume} {52}},\ \bibinfo {pages} {207} (\bibinfo {year} {2012})},\ \Eprint {http://arxiv.org/abs/https://onlinelibrary.wiley.com/doi/pdf/10.1002/ctpp.201100072} {https://onlinelibrary.wiley.com/doi/pdf/10.1002/ctpp.201100072} \BibitemShut {NoStop}%
\bibitem [{\citenamefont {Schwabe}\ \emph {et~al.}(2020)\citenamefont {Schwabe}, \citenamefont {Khrapak}, \citenamefont {Zhdanov}, \citenamefont {Pustylnik}, \citenamefont {Räth}, \citenamefont {Fink}, \citenamefont {Kretschmer}, \citenamefont {Lipaev}, \citenamefont {Molotkov}, \citenamefont {Schmitz}, \citenamefont {Thoma}, \citenamefont {Usachev}, \citenamefont {Zobnin}, \citenamefont {Padalka}, \citenamefont {Fortov}, \citenamefont {Petrov},\ and\ \citenamefont {Thomas}}]{Schwabe_2020}%
  \BibitemOpen
  \bibfield  {author} {\bibinfo {author} {\bibfnamefont {M.}~\bibnamefont {Schwabe}}, \bibinfo {author} {\bibfnamefont {S.~A.}\ \bibnamefont {Khrapak}}, \bibinfo {author} {\bibfnamefont {S.~K.}\ \bibnamefont {Zhdanov}}, \bibinfo {author} {\bibfnamefont {M.~Y.}\ \bibnamefont {Pustylnik}}, \bibinfo {author} {\bibfnamefont {C.}~\bibnamefont {Räth}}, \bibinfo {author} {\bibfnamefont {M.}~\bibnamefont {Fink}}, \bibinfo {author} {\bibfnamefont {M.}~\bibnamefont {Kretschmer}}, \bibinfo {author} {\bibfnamefont {A.~M.}\ \bibnamefont {Lipaev}}, \bibinfo {author} {\bibfnamefont {V.~I.}\ \bibnamefont {Molotkov}}, \bibinfo {author} {\bibfnamefont {A.~S.}\ \bibnamefont {Schmitz}}, \bibinfo {author} {\bibfnamefont {M.~H.}\ \bibnamefont {Thoma}}, \bibinfo {author} {\bibfnamefont {A.~D.}\ \bibnamefont {Usachev}}, \bibinfo {author} {\bibfnamefont {A.~V.}\ \bibnamefont {Zobnin}}, \bibinfo {author} {\bibfnamefont {G.~I.}\ \bibnamefont {Padalka}}, \bibinfo {author} {\bibfnamefont {V.~E.}\ \bibnamefont {Fortov}}, \bibinfo
  {author} {\bibfnamefont {O.~F.}\ \bibnamefont {Petrov}}, \ and\ \bibinfo {author} {\bibfnamefont {H.~M.}\ \bibnamefont {Thomas}},\ }\href {\doibase 10.1088/1367-2630/aba91b} {\bibfield  {journal} {\bibinfo  {journal} {New Journal of Physics}\ }\textbf {\bibinfo {volume} {22}},\ \bibinfo {pages} {083079} (\bibinfo {year} {2020})}\BibitemShut {NoStop}%
\bibitem [{\citenamefont {Jiang}\ \emph {et~al.}(2006)\citenamefont {Jiang}, \citenamefont {Hou}, \citenamefont {Wang},\ and\ \citenamefont {Mi\ifmmode \check{s}\else \v{s}\fi{}kovi\ifmmode~\acute{c}\else \'{c}\fi{}}}]{PhysRevE.73.016404}%
  \BibitemOpen
  \bibfield  {author} {\bibinfo {author} {\bibfnamefont {K.}~\bibnamefont {Jiang}}, \bibinfo {author} {\bibfnamefont {L.-J.}\ \bibnamefont {Hou}}, \bibinfo {author} {\bibfnamefont {Y.-N.}\ \bibnamefont {Wang}}, \ and\ \bibinfo {author} {\bibfnamefont {Z.~L.}\ \bibnamefont {Mi\ifmmode \check{s}\else \v{s}\fi{}kovi\ifmmode~\acute{c}\else \'{c}\fi{}}},\ }\href {\doibase 10.1103/PhysRevE.73.016404} {\bibfield  {journal} {\bibinfo  {journal} {Phys. Rev. E}\ }\textbf {\bibinfo {volume} {73}},\ \bibinfo {pages} {016404} (\bibinfo {year} {2006})}\BibitemShut {NoStop}%
\bibitem [{\citenamefont {Shukla}(2007)}]{Shukla_2007}%
  \BibitemOpen
  \bibfield  {author} {\bibinfo {author} {\bibfnamefont {P.~K.}\ \bibnamefont {Shukla}},\ }\href {\doibase 10.1088/0031-8949/76/5/N02} {\bibfield  {journal} {\bibinfo  {journal} {Physica Scripta}\ }\textbf {\bibinfo {volume} {76}},\ \bibinfo {pages} {C165} (\bibinfo {year} {2007})}\BibitemShut {NoStop}%
\bibitem [{\citenamefont {Kodanova}\ \emph {et~al.}(2017)\citenamefont {Kodanova}, \citenamefont {Issanova}, \citenamefont {Amirov}, \citenamefont {Ramazanov}, \citenamefont {Tikhonov},\ and\ \citenamefont {Moldabekov}}]{10.1016/j.mre.2017.07.005}%
  \BibitemOpen
  \bibfield  {author} {\bibinfo {author} {\bibfnamefont {S.}~\bibnamefont {Kodanova}}, \bibinfo {author} {\bibfnamefont {M.}~\bibnamefont {Issanova}}, \bibinfo {author} {\bibfnamefont {S.}~\bibnamefont {Amirov}}, \bibinfo {author} {\bibfnamefont {T.}~\bibnamefont {Ramazanov}}, \bibinfo {author} {\bibfnamefont {A.}~\bibnamefont {Tikhonov}}, \ and\ \bibinfo {author} {\bibfnamefont {Z.}~\bibnamefont {Moldabekov}},\ }\href {\doibase 10.1016/j.mre.2017.07.005} {\bibfield  {journal} {\bibinfo  {journal} {Matter and Radiation at Extremes}\ }\textbf {\bibinfo {volume} {3}},\ \bibinfo {pages} {40} (\bibinfo {year} {2017})},\ \Eprint {http://arxiv.org/abs/https://pubs.aip.org/aip/mre/article-pdf/3/1/40/16139000/40\_1\_online.pdf} {https://pubs.aip.org/aip/mre/article-pdf/3/1/40/16139000/40\_1\_online.pdf} \BibitemShut {NoStop}%
\bibitem [{\citenamefont {Issanova}\ \emph {et~al.}(2016)\citenamefont {Issanova}, \citenamefont {Kodanova}, \citenamefont {Ramazanov}, \citenamefont {Bastykova}, \citenamefont {Moldabekov},\ and\ \citenamefont {Meister}}]{Issanova_Meister_2016}%
  \BibitemOpen
  \bibfield  {author} {\bibinfo {author} {\bibfnamefont {M.~K.}\ \bibnamefont {Issanova}}, \bibinfo {author} {\bibfnamefont {S.~K.}\ \bibnamefont {Kodanova}}, \bibinfo {author} {\bibfnamefont {T.~S.}\ \bibnamefont {Ramazanov}}, \bibinfo {author} {\bibfnamefont {N.~K.}\ \bibnamefont {Bastykova}}, \bibinfo {author} {\bibfnamefont {Z.~A.}\ \bibnamefont {Moldabekov}}, \ and\ \bibinfo {author} {\bibfnamefont {C.-V.}\ \bibnamefont {Meister}},\ }\href {\doibase 10.1017/S026303461600032X} {\bibfield  {journal} {\bibinfo  {journal} {Laser and Particle Beams}\ }\textbf {\bibinfo {volume} {34}},\ \bibinfo {pages} {457–466} (\bibinfo {year} {2016})}\BibitemShut {NoStop}%
\bibitem [{\citenamefont {Ramazanov}\ \emph {et~al.}(2013)\citenamefont {Ramazanov}, \citenamefont {Kodanova}, \citenamefont {Moldabekov},\ and\ \citenamefont {Issanova}}]{10.1063/1.4829042}%
  \BibitemOpen
  \bibfield  {author} {\bibinfo {author} {\bibfnamefont {T.~S.}\ \bibnamefont {Ramazanov}}, \bibinfo {author} {\bibfnamefont {S.~K.}\ \bibnamefont {Kodanova}}, \bibinfo {author} {\bibfnamefont {Z.~A.}\ \bibnamefont {Moldabekov}}, \ and\ \bibinfo {author} {\bibfnamefont {M.~K.}\ \bibnamefont {Issanova}},\ }\href {\doibase 10.1063/1.4829042} {\bibfield  {journal} {\bibinfo  {journal} {Physics of Plasmas}\ }\textbf {\bibinfo {volume} {20}},\ \bibinfo {pages} {112702} (\bibinfo {year} {2013})},\ \Eprint {http://arxiv.org/abs/https://pubs.aip.org/aip/pop/article-pdf/doi/10.1063/1.4829042/14853668/112702\_1\_online.pdf} {https://pubs.aip.org/aip/pop/article-pdf/doi/10.1063/1.4829042/14853668/112702\_1\_online.pdf} \BibitemShut {NoStop}%
\bibitem [{\citenamefont {Hartmann}\ \emph {et~al.}(2013)\citenamefont {Hartmann}, \citenamefont {Donk\'o}, \citenamefont {Ott}, \citenamefont {K\"ahlert},\ and\ \citenamefont {Bonitz}}]{PhysRevLett.111.155002}%
  \BibitemOpen
  \bibfield  {author} {\bibinfo {author} {\bibfnamefont {P.}~\bibnamefont {Hartmann}}, \bibinfo {author} {\bibfnamefont {Z.}~\bibnamefont {Donk\'o}}, \bibinfo {author} {\bibfnamefont {T.}~\bibnamefont {Ott}}, \bibinfo {author} {\bibfnamefont {H.}~\bibnamefont {K\"ahlert}}, \ and\ \bibinfo {author} {\bibfnamefont {M.}~\bibnamefont {Bonitz}},\ }\href {\doibase 10.1103/PhysRevLett.111.155002} {\bibfield  {journal} {\bibinfo  {journal} {Phys. Rev. Lett.}\ }\textbf {\bibinfo {volume} {111}},\ \bibinfo {pages} {155002} (\bibinfo {year} {2013})}\BibitemShut {NoStop}%
\bibitem [{\citenamefont {Thomas}\ \emph {et~al.}(2015)\citenamefont {Thomas}, \citenamefont {Lynch}, \citenamefont {Konopka}, \citenamefont {Merlino},\ and\ \citenamefont {Rosenberg}}]{10.1063/1.4914089}%
  \BibitemOpen
  \bibfield  {author} {\bibinfo {author} {\bibfnamefont {J.}~\bibnamefont {Thomas}, \bibfnamefont {Edward}}, \bibinfo {author} {\bibfnamefont {B.}~\bibnamefont {Lynch}}, \bibinfo {author} {\bibfnamefont {U.}~\bibnamefont {Konopka}}, \bibinfo {author} {\bibfnamefont {R.~L.}\ \bibnamefont {Merlino}}, \ and\ \bibinfo {author} {\bibfnamefont {M.}~\bibnamefont {Rosenberg}},\ }\href {\doibase 10.1063/1.4914089} {\bibfield  {journal} {\bibinfo  {journal} {Physics of Plasmas}\ }\textbf {\bibinfo {volume} {22}},\ \bibinfo {pages} {030701} (\bibinfo {year} {2015})},\ \Eprint {http://arxiv.org/abs/https://pubs.aip.org/aip/pop/article-pdf/doi/10.1063/1.4914089/15897634/030701\_1\_online.pdf} {https://pubs.aip.org/aip/pop/article-pdf/doi/10.1063/1.4914089/15897634/030701\_1\_online.pdf} \BibitemShut {NoStop}%
\bibitem [{\citenamefont {Abdirakhmanov}\ \emph {et~al.}(2019)\citenamefont {Abdirakhmanov}, \citenamefont {Moldabekov}, \citenamefont {Kodanova}, \citenamefont {Dosbolayev},\ and\ \citenamefont {Ramazanov}}]{8688665}%
  \BibitemOpen
  \bibfield  {author} {\bibinfo {author} {\bibfnamefont {A.~R.}\ \bibnamefont {Abdirakhmanov}}, \bibinfo {author} {\bibfnamefont {Z.~A.}\ \bibnamefont {Moldabekov}}, \bibinfo {author} {\bibfnamefont {S.~K.}\ \bibnamefont {Kodanova}}, \bibinfo {author} {\bibfnamefont {M.~K.}\ \bibnamefont {Dosbolayev}}, \ and\ \bibinfo {author} {\bibfnamefont {T.~S.}\ \bibnamefont {Ramazanov}},\ }\href {\doibase 10.1109/TPS.2019.2906051} {\bibfield  {journal} {\bibinfo  {journal} {IEEE Transactions on Plasma Science}\ }\textbf {\bibinfo {volume} {47}},\ \bibinfo {pages} {3036} (\bibinfo {year} {2019})}\BibitemShut {NoStop}%
\bibitem [{\citenamefont {Ludwig}\ \emph {et~al.}(2018)\citenamefont {Ludwig}, \citenamefont {Jung}, \citenamefont {K{\"a}hlert}, \citenamefont {Joost}, \citenamefont {Greiner}, \citenamefont {Moldabekov}, \citenamefont {Carstensen}, \citenamefont {Sundar}, \citenamefont {Bonitz},\ and\ \citenamefont {Piel}}]{Ludwig2018}%
  \BibitemOpen
  \bibfield  {author} {\bibinfo {author} {\bibfnamefont {P.}~\bibnamefont {Ludwig}}, \bibinfo {author} {\bibfnamefont {H.}~\bibnamefont {Jung}}, \bibinfo {author} {\bibfnamefont {H.}~\bibnamefont {K{\"a}hlert}}, \bibinfo {author} {\bibfnamefont {J.-P.}\ \bibnamefont {Joost}}, \bibinfo {author} {\bibfnamefont {F.}~\bibnamefont {Greiner}}, \bibinfo {author} {\bibfnamefont {Z.}~\bibnamefont {Moldabekov}}, \bibinfo {author} {\bibfnamefont {J.}~\bibnamefont {Carstensen}}, \bibinfo {author} {\bibfnamefont {S.}~\bibnamefont {Sundar}}, \bibinfo {author} {\bibfnamefont {M.}~\bibnamefont {Bonitz}}, \ and\ \bibinfo {author} {\bibfnamefont {A.}~\bibnamefont {Piel}},\ }\href {\doibase 10.1140/epjd/e2017-80413-2} {\bibfield  {journal} {\bibinfo  {journal} {The European Physical Journal D}\ }\textbf {\bibinfo {volume} {72}},\ \bibinfo {pages} {82} (\bibinfo {year} {2018})}\BibitemShut {NoStop}%
\bibitem [{\citenamefont {Djienbekov}\ \emph {et~al.}(2024)\citenamefont {Djienbekov}, \citenamefont {Bastykova}, \citenamefont {Ramazanov},\ and\ \citenamefont {Kodanova}}]{Djienbekov2}%
  \BibitemOpen
  \bibfield  {author} {\bibinfo {author} {\bibfnamefont {N.~E.}\ \bibnamefont {Djienbekov}}, \bibinfo {author} {\bibfnamefont {N.~K.}\ \bibnamefont {Bastykova}}, \bibinfo {author} {\bibfnamefont {T.~S.}\ \bibnamefont {Ramazanov}}, \ and\ \bibinfo {author} {\bibfnamefont {S.~K.}\ \bibnamefont {Kodanova}},\ }\href {\doibase 10.1038/s41598-024-64866-z} {\bibfield  {journal} {\bibinfo  {journal} {Scientific Reports}\ }\textbf {\bibinfo {volume} {14}},\ \bibinfo {pages} {15042} (\bibinfo {year} {2024})}\BibitemShut {NoStop}%
\bibitem [{\citenamefont {Liu}\ and\ \citenamefont {Chew}(2007)}]{Liu_2007}%
  \BibitemOpen
  \bibfield  {author} {\bibinfo {author} {\bibfnamefont {Y.}~\bibnamefont {Liu}}\ and\ \bibinfo {author} {\bibfnamefont {L.~Y.}\ \bibnamefont {Chew}},\ }\href {\doibase 10.1088/1751-8113/40/33/027} {\bibfield  {journal} {\bibinfo  {journal} {Journal of Physics A: Mathematical and Theoretical}\ }\textbf {\bibinfo {volume} {40}},\ \bibinfo {pages} {10383} (\bibinfo {year} {2007})}\BibitemShut {NoStop}%
\bibitem [{\citenamefont {Yang}\ \emph {et~al.}(2020)\citenamefont {Yang}, \citenamefont {Kong}, \citenamefont {Liu},\ and\ \citenamefont {Li}}]{yang2020dynamics}%
  \BibitemOpen
  \bibfield  {author} {\bibinfo {author} {\bibfnamefont {F.}~\bibnamefont {Yang}}, \bibinfo {author} {\bibfnamefont {W.}~\bibnamefont {Kong}}, \bibinfo {author} {\bibfnamefont {S.}~\bibnamefont {Liu}}, \ and\ \bibinfo {author} {\bibfnamefont {Y.}~\bibnamefont {Li}},\ }\href@noop {} {\bibfield  {journal} {\bibinfo  {journal} {Physics of Plasmas}\ }\textbf {\bibinfo {volume} {27}} (\bibinfo {year} {2020})}\BibitemShut {NoStop}%
\bibitem [{\citenamefont {Lapenta}(1995)}]{PhysRevLett.75.4409}%
  \BibitemOpen
  \bibfield  {author} {\bibinfo {author} {\bibfnamefont {G.}~\bibnamefont {Lapenta}},\ }\href {\doibase 10.1103/PhysRevLett.75.4409} {\bibfield  {journal} {\bibinfo  {journal} {Phys. Rev. Lett.}\ }\textbf {\bibinfo {volume} {75}},\ \bibinfo {pages} {4409} (\bibinfo {year} {1995})}\BibitemShut {NoStop}%
\bibitem [{\citenamefont {Sukhinin}\ and\ \citenamefont {Fedoseev}(2010)}]{5510173}%
  \BibitemOpen
  \bibfield  {author} {\bibinfo {author} {\bibfnamefont {G.~I.}\ \bibnamefont {Sukhinin}}\ and\ \bibinfo {author} {\bibfnamefont {A.~V.}\ \bibnamefont {Fedoseev}},\ }\href {\doibase 10.1109/TPS.2010.2052370} {\bibfield  {journal} {\bibinfo  {journal} {IEEE Transactions on Plasma Science}\ }\textbf {\bibinfo {volume} {38}},\ \bibinfo {pages} {2345} (\bibinfo {year} {2010})}\BibitemShut {NoStop}%
\bibitem [{\citenamefont {Ramazanov}\ \emph {et~al.}(2016)\citenamefont {Ramazanov}, \citenamefont {Moldabekov},\ and\ \citenamefont {Gabdullin}}]{PhysRevE.93.053204}%
  \BibitemOpen
  \bibfield  {author} {\bibinfo {author} {\bibfnamefont {T.~S.}\ \bibnamefont {Ramazanov}}, \bibinfo {author} {\bibfnamefont {Z.~A.}\ \bibnamefont {Moldabekov}}, \ and\ \bibinfo {author} {\bibfnamefont {M.~T.}\ \bibnamefont {Gabdullin}},\ }\href {\doibase 10.1103/PhysRevE.93.053204} {\bibfield  {journal} {\bibinfo  {journal} {Phys. Rev. E}\ }\textbf {\bibinfo {volume} {93}},\ \bibinfo {pages} {053204} (\bibinfo {year} {2016})}\BibitemShut {NoStop}%
\bibitem [{\citenamefont {Aldakul}\ \emph {et~al.}(2020)\citenamefont {Aldakul}, \citenamefont {Moldabekov},\ and\ \citenamefont {Ramazanov}}]{PhysRevE.102.033205}%
  \BibitemOpen
  \bibfield  {author} {\bibinfo {author} {\bibfnamefont {Y.~K.}\ \bibnamefont {Aldakul}}, \bibinfo {author} {\bibfnamefont {Z.~A.}\ \bibnamefont {Moldabekov}}, \ and\ \bibinfo {author} {\bibfnamefont {T.~S.}\ \bibnamefont {Ramazanov}},\ }\href {\doibase 10.1103/PhysRevE.102.033205} {\bibfield  {journal} {\bibinfo  {journal} {Phys. Rev. E}\ }\textbf {\bibinfo {volume} {102}},\ \bibinfo {pages} {033205} (\bibinfo {year} {2020})}\BibitemShut {NoStop}%
\bibitem [{\citenamefont {Boudec}\ \emph {et~al.}(2024)\citenamefont {Boudec}, \citenamefont {Oregel-Chaumont}, \citenamefont {Rachidi}, \citenamefont {Rubinstein},\ and\ \citenamefont {Vega}}]{boudec2024cartesiansphericalmultipoleexpansions}%
  \BibitemOpen
  \bibfield  {author} {\bibinfo {author} {\bibfnamefont {E.~L.}\ \bibnamefont {Boudec}}, \bibinfo {author} {\bibfnamefont {T.}~\bibnamefont {Oregel-Chaumont}}, \bibinfo {author} {\bibfnamefont {F.}~\bibnamefont {Rachidi}}, \bibinfo {author} {\bibfnamefont {M.}~\bibnamefont {Rubinstein}}, \ and\ \bibinfo {author} {\bibfnamefont {F.}~\bibnamefont {Vega}},\ }\href {https://arxiv.org/abs/2408.12303} {\enquote {\bibinfo {title} {Cartesian and spherical multipole expansions in anisotropic media},}\ } (\bibinfo {year} {2024}),\ \Eprint {http://arxiv.org/abs/2408.12303} {arXiv:2408.12303 [physics.optics]} \BibitemShut {NoStop}%
\bibitem [{\citenamefont {Davletov}\ \emph {et~al.}(2014)\citenamefont {Davletov}, \citenamefont {Yerimbetova}, \citenamefont {Mukhametkarimov},\ and\ \citenamefont {Ospanova}}]{10.1063/1.4887009}%
  \BibitemOpen
  \bibfield  {author} {\bibinfo {author} {\bibfnamefont {A.~E.}\ \bibnamefont {Davletov}}, \bibinfo {author} {\bibfnamefont {L.~T.}\ \bibnamefont {Yerimbetova}}, \bibinfo {author} {\bibfnamefont {Y.~S.}\ \bibnamefont {Mukhametkarimov}}, \ and\ \bibinfo {author} {\bibfnamefont {A.~K.}\ \bibnamefont {Ospanova}},\ }\href {\doibase 10.1063/1.4887009} {\bibfield  {journal} {\bibinfo  {journal} {Physics of Plasmas}\ }\textbf {\bibinfo {volume} {21}},\ \bibinfo {pages} {073704} (\bibinfo {year} {2014})},\ \Eprint {http://arxiv.org/abs/https://pubs.aip.org/aip/pop/article-pdf/doi/10.1063/1.4887009/15696447/073704\_1\_online.pdf} {https://pubs.aip.org/aip/pop/article-pdf/doi/10.1063/1.4887009/15696447/073704\_1\_online.pdf} \BibitemShut {NoStop}%
\bibitem [{\citenamefont {Aldakulov}\ \emph {et~al.}(2020)\citenamefont {Aldakulov}, \citenamefont {Moldabekov}, \citenamefont {Muratov},\ and\ \citenamefont {Ramazanov}}]{Aldakulov_2020}%
  \BibitemOpen
  \bibfield  {author} {\bibinfo {author} {\bibfnamefont {Y.~Q.}\ \bibnamefont {Aldakulov}}, \bibinfo {author} {\bibfnamefont {Z.~A.}\ \bibnamefont {Moldabekov}}, \bibinfo {author} {\bibfnamefont {M.}~\bibnamefont {Muratov}}, \ and\ \bibinfo {author} {\bibfnamefont {T.~S.}\ \bibnamefont {Ramazanov}},\ }\href {\doibase 10.7567/1347-4065/ab6565} {\bibfield  {journal} {\bibinfo  {journal} {Japanese Journal of Applied Physics}\ }\textbf {\bibinfo {volume} {59}},\ \bibinfo {pages} {SHHE02} (\bibinfo {year} {2020})}\BibitemShut {NoStop}%
\bibitem [{\citenamefont {Sundar}\ and\ \citenamefont {Moldabekov}(2020)}]{Sundar_2020}%
  \BibitemOpen
  \bibfield  {author} {\bibinfo {author} {\bibfnamefont {S.}~\bibnamefont {Sundar}}\ and\ \bibinfo {author} {\bibfnamefont {Z.~A.}\ \bibnamefont {Moldabekov}},\ }\href {\doibase 10.1088/1367-2630/ab7bd2} {\bibfield  {journal} {\bibinfo  {journal} {New Journal of Physics}\ }\textbf {\bibinfo {volume} {22}},\ \bibinfo {pages} {033028} (\bibinfo {year} {2020})}\BibitemShut {NoStop}%
\bibitem [{\citenamefont {Sundar}\ and\ \citenamefont {Moldabekov}(2019)}]{PhysRevE.99.063202}%
  \BibitemOpen
  \bibfield  {author} {\bibinfo {author} {\bibfnamefont {S.}~\bibnamefont {Sundar}}\ and\ \bibinfo {author} {\bibfnamefont {Z.~A.}\ \bibnamefont {Moldabekov}},\ }\href {\doibase 10.1103/PhysRevE.99.063202} {\bibfield  {journal} {\bibinfo  {journal} {Phys. Rev. E}\ }\textbf {\bibinfo {volume} {99}},\ \bibinfo {pages} {063202} (\bibinfo {year} {2019})}\BibitemShut {NoStop}%
\bibitem [{\citenamefont {Vermant}\ and\ \citenamefont {Solomon}(2005)}]{Vermant_2005}%
  \BibitemOpen
  \bibfield  {author} {\bibinfo {author} {\bibfnamefont {J.}~\bibnamefont {Vermant}}\ and\ \bibinfo {author} {\bibfnamefont {M.~J.}\ \bibnamefont {Solomon}},\ }\href {\doibase 10.1088/0953-8984/17/4/R02} {\bibfield  {journal} {\bibinfo  {journal} {Journal of Physics: Condensed Matter}\ }\textbf {\bibinfo {volume} {17}},\ \bibinfo {pages} {R187} (\bibinfo {year} {2005})}\BibitemShut {NoStop}%
\bibitem [{\citenamefont {Royall}\ \emph {et~al.}(2024)\citenamefont {Royall}, \citenamefont {Charbonneau}, \citenamefont {Dijkstra}, \citenamefont {Russo}, \citenamefont {Smallenburg}, \citenamefont {Speck},\ and\ \citenamefont {Valeriani}}]{RevModPhys.96.045003}%
  \BibitemOpen
  \bibfield  {author} {\bibinfo {author} {\bibfnamefont {C.~P.}\ \bibnamefont {Royall}}, \bibinfo {author} {\bibfnamefont {P.}~\bibnamefont {Charbonneau}}, \bibinfo {author} {\bibfnamefont {M.}~\bibnamefont {Dijkstra}}, \bibinfo {author} {\bibfnamefont {J.}~\bibnamefont {Russo}}, \bibinfo {author} {\bibfnamefont {F.}~\bibnamefont {Smallenburg}}, \bibinfo {author} {\bibfnamefont {T.}~\bibnamefont {Speck}}, \ and\ \bibinfo {author} {\bibfnamefont {C.}~\bibnamefont {Valeriani}},\ }\href {\doibase 10.1103/RevModPhys.96.045003} {\bibfield  {journal} {\bibinfo  {journal} {Rev. Mod. Phys.}\ }\textbf {\bibinfo {volume} {96}},\ \bibinfo {pages} {045003} (\bibinfo {year} {2024})}\BibitemShut {NoStop}%
\bibitem [{\citenamefont {Anderson}\ and\ \citenamefont {Lekkerkerker}(2002)}]{Anderson2002}%
  \BibitemOpen
  \bibfield  {author} {\bibinfo {author} {\bibfnamefont {V.~J.}\ \bibnamefont {Anderson}}\ and\ \bibinfo {author} {\bibfnamefont {H.~N.~W.}\ \bibnamefont {Lekkerkerker}},\ }\href {\doibase 10.1038/416811a} {\bibfield  {journal} {\bibinfo  {journal} {Nature}\ }\textbf {\bibinfo {volume} {416}},\ \bibinfo {pages} {811} (\bibinfo {year} {2002})}\BibitemShut {NoStop}%
\bibitem [{\citenamefont {Bonitz}\ \emph {et~al.}(2010{\natexlab{a}})\citenamefont {Bonitz}, \citenamefont {Henning},\ and\ \citenamefont {Block}}]{Bonitz_2010}%
  \BibitemOpen
  \bibfield  {author} {\bibinfo {author} {\bibfnamefont {M.}~\bibnamefont {Bonitz}}, \bibinfo {author} {\bibfnamefont {C.}~\bibnamefont {Henning}}, \ and\ \bibinfo {author} {\bibfnamefont {D.}~\bibnamefont {Block}},\ }\href {\doibase 10.1088/0034-4885/73/6/066501} {\bibfield  {journal} {\bibinfo  {journal} {Reports on Progress in Physics}\ }\textbf {\bibinfo {volume} {73}},\ \bibinfo {pages} {066501} (\bibinfo {year} {2010}{\natexlab{a}})}\BibitemShut {NoStop}%
\bibitem [{\citenamefont {Malijevsk{\'y}}\ and\ \citenamefont {Kolafa}(2008)}]{Malijevský2008}%
  \BibitemOpen
  \bibfield  {author} {\bibinfo {author} {\bibfnamefont {A.}~\bibnamefont {Malijevsk{\'y}}}\ and\ \bibinfo {author} {\bibfnamefont {J.}~\bibnamefont {Kolafa}},\ }\enquote {\bibinfo {title} {Structure of hard spheres and related systems},}\ in\ \href {\doibase 10.1007/978-3-540-78767-9_1} {\emph {\bibinfo {booktitle} {Theory and Simulation of Hard-Sphere Fluids and Related Systems}}},\ \bibinfo {editor} {edited by\ \bibinfo {editor} {\bibfnamefont {{\'A}.}~\bibnamefont {Mulero}}}\ (\bibinfo  {publisher} {Springer Berlin Heidelberg},\ \bibinfo {address} {Berlin, Heidelberg},\ \bibinfo {year} {2008})\ pp.\ \bibinfo {pages} {1--26}\BibitemShut {NoStop}%
\bibitem [{\citenamefont {Moldabekov}\ \emph {et~al.}(2021)\citenamefont {Moldabekov}, \citenamefont {Aldakul}, \citenamefont {Bastykova}, \citenamefont {Sundar},\ and\ \citenamefont {Cangi}}]{PhysRevResearch.3.043187}%
  \BibitemOpen
  \bibfield  {author} {\bibinfo {author} {\bibfnamefont {Z.~A.}\ \bibnamefont {Moldabekov}}, \bibinfo {author} {\bibfnamefont {Y.~K.}\ \bibnamefont {Aldakul}}, \bibinfo {author} {\bibfnamefont {N.~K.}\ \bibnamefont {Bastykova}}, \bibinfo {author} {\bibfnamefont {S.}~\bibnamefont {Sundar}}, \ and\ \bibinfo {author} {\bibfnamefont {A.}~\bibnamefont {Cangi}},\ }\href {\doibase 10.1103/PhysRevResearch.3.043187} {\bibfield  {journal} {\bibinfo  {journal} {Phys. Rev. Res.}\ }\textbf {\bibinfo {volume} {3}},\ \bibinfo {pages} {043187} (\bibinfo {year} {2021})}\BibitemShut {NoStop}%
\bibitem [{\citenamefont {Tsytovich}\ and\ \citenamefont {Gusein-zade}(2013)}]{Tsytovich2013}%
  \BibitemOpen
  \bibfield  {author} {\bibinfo {author} {\bibfnamefont {V.~N.}\ \bibnamefont {Tsytovich}}\ and\ \bibinfo {author} {\bibfnamefont {N.~G.}\ \bibnamefont {Gusein-zade}},\ }\href {\doibase 10.1134/S1063780X13070088} {\bibfield  {journal} {\bibinfo  {journal} {Plasma Physics Reports}\ }\textbf {\bibinfo {volume} {39}},\ \bibinfo {pages} {515} (\bibinfo {year} {2013})}\BibitemShut {NoStop}%
\bibitem [{\citenamefont {Zoetekouw}\ and\ \citenamefont {van Roij}(2006)}]{PhysRevLett.97.258302}%
  \BibitemOpen
  \bibfield  {author} {\bibinfo {author} {\bibfnamefont {B.}~\bibnamefont {Zoetekouw}}\ and\ \bibinfo {author} {\bibfnamefont {R.}~\bibnamefont {van Roij}},\ }\href {\doibase 10.1103/PhysRevLett.97.258302} {\bibfield  {journal} {\bibinfo  {journal} {Phys. Rev. Lett.}\ }\textbf {\bibinfo {volume} {97}},\ \bibinfo {pages} {258302} (\bibinfo {year} {2006})}\BibitemShut {NoStop}%
\bibitem [{\citenamefont {Spreiter}\ and\ \citenamefont {Walter}(1999)}]{Spreiter1999ClassicalMD}%
  \BibitemOpen
  \bibfield  {author} {\bibinfo {author} {\bibfnamefont {Q.}~\bibnamefont {Spreiter}}\ and\ \bibinfo {author} {\bibfnamefont {M.}~\bibnamefont {Walter}},\ }\href@noop {} {\bibfield  {journal} {\bibinfo  {journal} {Journal of Computational Physics}\ }\textbf {\bibinfo {volume} {152}},\ \bibinfo {pages} {102} (\bibinfo {year} {1999})}\BibitemShut {NoStop}%
\bibitem [{\citenamefont {M\"uller-Plathe}(1999)}]{Muller}%
  \BibitemOpen
  \bibfield  {author} {\bibinfo {author} {\bibfnamefont {F.}~\bibnamefont {M\"uller-Plathe}},\ }\href {\doibase 10.1103/PhysRevE.59.4894} {\bibfield  {journal} {\bibinfo  {journal} {Phys. Rev. E}\ }\textbf {\bibinfo {volume} {59}},\ \bibinfo {pages} {4894} (\bibinfo {year} {1999})}\BibitemShut {NoStop}%
\bibitem [{\citenamefont {Sanbonmatsu}\ and\ \citenamefont {Murillo}(2001)}]{Sanbonmatsu}%
  \BibitemOpen
  \bibfield  {author} {\bibinfo {author} {\bibfnamefont {K.~Y.}\ \bibnamefont {Sanbonmatsu}}\ and\ \bibinfo {author} {\bibfnamefont {M.~S.}\ \bibnamefont {Murillo}},\ }\href {\doibase 10.1103/PhysRevLett.86.1215} {\bibfield  {journal} {\bibinfo  {journal} {Phys. Rev. Lett.}\ }\textbf {\bibinfo {volume} {86}},\ \bibinfo {pages} {1215} (\bibinfo {year} {2001})}\BibitemShut {NoStop}%
\bibitem [{\citenamefont {Donk\'o}\ \emph {et~al.}(2006)\citenamefont {Donk\'o}, \citenamefont {Goree}, \citenamefont {Hartmann},\ and\ \citenamefont {Kutasi}}]{Hartmann}%
  \BibitemOpen
  \bibfield  {author} {\bibinfo {author} {\bibfnamefont {Z.}~\bibnamefont {Donk\'o}}, \bibinfo {author} {\bibfnamefont {J.}~\bibnamefont {Goree}}, \bibinfo {author} {\bibfnamefont {P.}~\bibnamefont {Hartmann}}, \ and\ \bibinfo {author} {\bibfnamefont {K.}~\bibnamefont {Kutasi}},\ }\href {\doibase 10.1103/PhysRevLett.96.145003} {\bibfield  {journal} {\bibinfo  {journal} {Phys. Rev. Lett.}\ }\textbf {\bibinfo {volume} {96}},\ \bibinfo {pages} {145003} (\bibinfo {year} {2006})}\BibitemShut {NoStop}%
\bibitem [{\citenamefont {Djienbekov}\ \emph {et~al.}(2022{\natexlab{b}})\citenamefont {Djienbekov}, \citenamefont {Bastykova}, \citenamefont {Bekbussyn}, \citenamefont {Ramazanov},\ and\ \citenamefont {Kodanova}}]{Djienbekov}%
  \BibitemOpen
  \bibfield  {author} {\bibinfo {author} {\bibfnamefont {N.~E.}\ \bibnamefont {Djienbekov}}, \bibinfo {author} {\bibfnamefont {N.~K.}\ \bibnamefont {Bastykova}}, \bibinfo {author} {\bibfnamefont {A.~M.}\ \bibnamefont {Bekbussyn}}, \bibinfo {author} {\bibfnamefont {T.~S.}\ \bibnamefont {Ramazanov}}, \ and\ \bibinfo {author} {\bibfnamefont {S.~K.}\ \bibnamefont {Kodanova}},\ }\href {\doibase 10.1103/PhysRevE.106.065203} {\bibfield  {journal} {\bibinfo  {journal} {Phys. Rev. E}\ }\textbf {\bibinfo {volume} {106}},\ \bibinfo {pages} {065203} (\bibinfo {year} {2022}{\natexlab{b}})}\BibitemShut {NoStop}%
\bibitem [{\citenamefont {Donk\'o}\ \emph {et~al.}(2009)\citenamefont {Donk\'o}, \citenamefont {Goree}, \citenamefont {Hartmann},\ and\ \citenamefont {Liu}}]{Donko2}%
  \BibitemOpen
  \bibfield  {author} {\bibinfo {author} {\bibfnamefont {Z.}~\bibnamefont {Donk\'o}}, \bibinfo {author} {\bibfnamefont {J.}~\bibnamefont {Goree}}, \bibinfo {author} {\bibfnamefont {P.}~\bibnamefont {Hartmann}}, \ and\ \bibinfo {author} {\bibfnamefont {B.}~\bibnamefont {Liu}},\ }\href {\doibase 10.1103/PhysRevE.79.026401} {\bibfield  {journal} {\bibinfo  {journal} {Phys. Rev. E}\ }\textbf {\bibinfo {volume} {79}},\ \bibinfo {pages} {026401} (\bibinfo {year} {2009})}\BibitemShut {NoStop}%
\bibitem [{\citenamefont {Baalrud}\ and\ \citenamefont {Daligault}(2014)}]{Baalrud_pop_2014}%
  \BibitemOpen
  \bibfield  {author} {\bibinfo {author} {\bibfnamefont {S.~D.}\ \bibnamefont {Baalrud}}\ and\ \bibinfo {author} {\bibfnamefont {J.}~\bibnamefont {Daligault}},\ }\href {\doibase 10.1063/1.4875282} {\bibfield  {journal} {\bibinfo  {journal} {Physics of Plasmas}\ }\textbf {\bibinfo {volume} {21}},\ \bibinfo {pages} {055707} (\bibinfo {year} {2014})},\ \Eprint {http://arxiv.org/abs/https://pubs.aip.org/aip/pop/article-pdf/doi/10.1063/1.4875282/15921503/055707\_1\_online.pdf} {https://pubs.aip.org/aip/pop/article-pdf/doi/10.1063/1.4875282/15921503/055707\_1\_online.pdf} \BibitemShut {NoStop}%
\bibitem [{\citenamefont {Baalrud}\ and\ \citenamefont {Daligault}(2013)}]{Baalrud_prl_2013}%
  \BibitemOpen
  \bibfield  {author} {\bibinfo {author} {\bibfnamefont {S.~D.}\ \bibnamefont {Baalrud}}\ and\ \bibinfo {author} {\bibfnamefont {J.}~\bibnamefont {Daligault}},\ }\href {\doibase 10.1103/PhysRevLett.110.235001} {\bibfield  {journal} {\bibinfo  {journal} {Phys. Rev. Lett.}\ }\textbf {\bibinfo {volume} {110}},\ \bibinfo {pages} {235001} (\bibinfo {year} {2013})}\BibitemShut {NoStop}%
\bibitem [{\citenamefont {Müller-Plathe}(1997)}]{Müller}%
  \BibitemOpen
  \bibfield  {author} {\bibinfo {author} {\bibfnamefont {F.}~\bibnamefont {Müller-Plathe}},\ }\href {\doibase 10.1063/1.473271} {\bibfield  {journal} {\bibinfo  {journal} {The Journal of Chemical Physics}\ }\textbf {\bibinfo {volume} {106}},\ \bibinfo {pages} {6082} (\bibinfo {year} {1997})},\ \Eprint {http://arxiv.org/abs/https://pubs.aip.org/aip/jcp/article-pdf/106/14/6082/10782785/6082\_1\_online.pdf} {https://pubs.aip.org/aip/jcp/article-pdf/106/14/6082/10782785/6082\_1\_online.pdf} \BibitemShut {NoStop}%
\bibitem [{\citenamefont {Matthews}\ \emph {et~al.}(2016)\citenamefont {Matthews}, \citenamefont {Coleman},\ and\ \citenamefont {Hyde}}]{7302071}%
  \BibitemOpen
  \bibfield  {author} {\bibinfo {author} {\bibfnamefont {L.~S.}\ \bibnamefont {Matthews}}, \bibinfo {author} {\bibfnamefont {D.~A.}\ \bibnamefont {Coleman}}, \ and\ \bibinfo {author} {\bibfnamefont {T.~W.}\ \bibnamefont {Hyde}},\ }\href {\doibase 10.1109/TPS.2015.2485162} {\bibfield  {journal} {\bibinfo  {journal} {IEEE Transactions on Plasma Science}\ }\textbf {\bibinfo {volume} {44}},\ \bibinfo {pages} {519} (\bibinfo {year} {2016})}\BibitemShut {NoStop}%
\bibitem [{\citenamefont {Schweigert}\ \emph {et~al.}(1999)\citenamefont {Schweigert}, \citenamefont {Schweigert},\ and\ \citenamefont {Peeters}}]{PhysRevB.60.14665}%
  \BibitemOpen
  \bibfield  {author} {\bibinfo {author} {\bibfnamefont {I.~V.}\ \bibnamefont {Schweigert}}, \bibinfo {author} {\bibfnamefont {V.~A.}\ \bibnamefont {Schweigert}}, \ and\ \bibinfo {author} {\bibfnamefont {F.~M.}\ \bibnamefont {Peeters}},\ }\href {\doibase 10.1103/PhysRevB.60.14665} {\bibfield  {journal} {\bibinfo  {journal} {Phys. Rev. B}\ }\textbf {\bibinfo {volume} {60}},\ \bibinfo {pages} {14665} (\bibinfo {year} {1999})}\BibitemShut {NoStop}%
\bibitem [{\citenamefont {Radzvilavi\ifmmode~\check{c}\else \v{c}\fi{}ius}(2012)}]{PhysRevE.86.051111}%
  \BibitemOpen
  \bibfield  {author} {\bibinfo {author} {\bibfnamefont {A.~b.~u.}\ \bibnamefont {Radzvilavi\ifmmode~\check{c}\else \v{c}\fi{}ius}},\ }\href {\doibase 10.1103/PhysRevE.86.051111} {\bibfield  {journal} {\bibinfo  {journal} {Phys. Rev. E}\ }\textbf {\bibinfo {volume} {86}},\ \bibinfo {pages} {051111} (\bibinfo {year} {2012})}\BibitemShut {NoStop}%
\bibitem [{\citenamefont {Ott}\ and\ \citenamefont {Bonitz}(2009)}]{PhysRevLett.103.195001}%
  \BibitemOpen
  \bibfield  {author} {\bibinfo {author} {\bibfnamefont {T.}~\bibnamefont {Ott}}\ and\ \bibinfo {author} {\bibfnamefont {M.}~\bibnamefont {Bonitz}},\ }\href {\doibase 10.1103/PhysRevLett.103.195001} {\bibfield  {journal} {\bibinfo  {journal} {Phys. Rev. Lett.}\ }\textbf {\bibinfo {volume} {103}},\ \bibinfo {pages} {195001} (\bibinfo {year} {2009})}\BibitemShut {NoStop}%
\bibitem [{\citenamefont {Bonitz}\ \emph {et~al.}(2010{\natexlab{b}})\citenamefont {Bonitz}, \citenamefont {Donk\'o}, \citenamefont {Ott}, \citenamefont {K\"ahlert},\ and\ \citenamefont {Hartmann}}]{Bonitz}%
  \BibitemOpen
  \bibfield  {author} {\bibinfo {author} {\bibfnamefont {M.}~\bibnamefont {Bonitz}}, \bibinfo {author} {\bibfnamefont {Z.}~\bibnamefont {Donk\'o}}, \bibinfo {author} {\bibfnamefont {T.}~\bibnamefont {Ott}}, \bibinfo {author} {\bibfnamefont {H.}~\bibnamefont {K\"ahlert}}, \ and\ \bibinfo {author} {\bibfnamefont {P.}~\bibnamefont {Hartmann}},\ }\href {\doibase 10.1103/PhysRevLett.105.055002} {\bibfield  {journal} {\bibinfo  {journal} {Phys. Rev. Lett.}\ }\textbf {\bibinfo {volume} {105}},\ \bibinfo {pages} {055002} (\bibinfo {year} {2010}{\natexlab{b}})}\BibitemShut {NoStop}%
\bibitem [{\citenamefont {Hartmann}\ \emph {et~al.}(2019)\citenamefont {Hartmann}, \citenamefont {Reyes}, \citenamefont {Kostadinova}, \citenamefont {Matthews}, \citenamefont {Hyde}, \citenamefont {Masheyeva}, \citenamefont {Dzhumagulova}, \citenamefont {Ramazanov}, \citenamefont {Ott}, \citenamefont {K\"ahlert}, \citenamefont {Bonitz}, \citenamefont {Korolov},\ and\ \citenamefont {Donk\'o}}]{PhysRevE.99.013203}%
  \BibitemOpen
  \bibfield  {author} {\bibinfo {author} {\bibfnamefont {P.}~\bibnamefont {Hartmann}}, \bibinfo {author} {\bibfnamefont {J.~C.}\ \bibnamefont {Reyes}}, \bibinfo {author} {\bibfnamefont {E.~G.}\ \bibnamefont {Kostadinova}}, \bibinfo {author} {\bibfnamefont {L.~S.}\ \bibnamefont {Matthews}}, \bibinfo {author} {\bibfnamefont {T.~W.}\ \bibnamefont {Hyde}}, \bibinfo {author} {\bibfnamefont {R.~U.}\ \bibnamefont {Masheyeva}}, \bibinfo {author} {\bibfnamefont {K.~N.}\ \bibnamefont {Dzhumagulova}}, \bibinfo {author} {\bibfnamefont {T.~S.}\ \bibnamefont {Ramazanov}}, \bibinfo {author} {\bibfnamefont {T.}~\bibnamefont {Ott}}, \bibinfo {author} {\bibfnamefont {H.}~\bibnamefont {K\"ahlert}}, \bibinfo {author} {\bibfnamefont {M.}~\bibnamefont {Bonitz}}, \bibinfo {author} {\bibfnamefont {I.}~\bibnamefont {Korolov}}, \ and\ \bibinfo {author} {\bibfnamefont {Z.}~\bibnamefont {Donk\'o}},\ }\href {\doibase 10.1103/PhysRevE.99.013203} {\bibfield  {journal} {\bibinfo  {journal} {Phys. Rev. E}\ }\textbf {\bibinfo {volume} {99}},\
  \bibinfo {pages} {013203} (\bibinfo {year} {2019})}\BibitemShut {NoStop}%
\end{thebibliography}%
\end{document}